\documentclass[pdflatex]{sn-jnl}

\usepackage{graphicx}%
\usepackage{multirow}%
\usepackage{amsmath,amssymb,amsfonts}%
\usepackage{amsthm}%
\usepackage{mathrsfs}%
\usepackage[title]{appendix}%
\usepackage{xcolor}%
\usepackage{textcomp}%
\usepackage{manyfoot}%
\usepackage{booktabs}%
\usepackage{algorithm}%
\usepackage{algorithmicx}%
\usepackage{algpseudocode}%
\usepackage{listings}%
\usepackage{enumitem}
\usepackage{capt-of}
\usepackage{adjustbox}
\usepackage{pdfpages}
\theoremstyle{thmstyleone}

\usepackage[
    backend=biber,
    style=apa,
    natbib=true
]{biblatex}

\theoremstyle{thmstyletwo}%

\theoremstyle{thmstylethree}%

\definecolor{bblack}{HTML}{363033}
\definecolor{dblau}{RGB}{21,50,104}
\definecolor{hblau}{RGB}{188,206,226}
\definecolor{hhblau}{RGB}{0,145,200}
\definecolor{chamois}{RGB}{243,226,216}
\definecolor{altweis}{RGB}{246,244,240}
\definecolor{dgrau}{RGB}{59,59,58}
\definecolor{hgrau}{RGB}{135,135,134}
\usepackage{tikz}

\begin{document}

\title[AI-supported worked examples in physics education]{AI-generated support, worked examples, and script-based learning in undergraduate physics: A comparative study}

\author*[1]{\fnm{Larissa} \sur{Hahn}}\email{larissa.hahn@uni-goettingen.de}

\author[1]{\fnm{Marie Aylin} \sur{Falkenberg}}\email{aylin.falkenberg@uni-goettingen.de}

\author[1]{\fnm{Marten} \sur{Hogelücht}}\email{marten.hogeluecht@uni-goettingen.de}

\author[1]{\fnm{Pascal} \sur{Klein}}\email{pascal.klein@uni-goettingen.de}

\affil[1]{\orgdiv{Physics Education Research}, \orgname{University of Göttingen}, \orgaddress{\street{Friedrich-Hund-Platz 1}, \city{Göttingen}, \postcode{37077},  \country{Germany}}}

\abstract{Generative AI is increasingly used to support students during learning and problem solving, but its instructional value depends on how it is embedded in established learning approaches. Worked examples guide learners through the steps and reasoning of a solution; conversational AI may extend them by responding to individual questions. To isolate this added value, we compared a chatbot-delivered intelligent worked example with a matched static worked example and a lecture script. Nine recitation groups in an undergraduate electrodynamics course were cluster-randomized to the three conditions, and 86 students worked individually on a Gauss’s law problem. Concept-test accuracy and self-efficacy increased significantly in all conditions, but changes did not differ significantly between groups ($p=.07$ for concept-test accuracy and $p=.38$ for self-efficacy). Four weeks later, in a reduced sample with selective participation and pre-existing group differences, students in the static worked-example condition scored higher than those in the script condition ($p=.004$, $d=1.22$), whereas the two worked-example conditions did not differ significantly. Usability, but not perceived educational value, was higher for the intelligent worked example than for the script ($p=.011$, $d=0.88$). In a subsequent homework assignment on Ampère’s law with free choice of learning material ($N=46$), $62-71\%$ of students in each condition reported using only the static worked example, at most $8\%$ used exclusively the intelligent one, and $15-31\%$ used neither. Thus, conversational delivery showed no clear additional learning benefit under the present conditions. More broadly, a learning format that is rated favorably need not be the one students voluntarily adopt.}


\maketitle

\section{Introduction}\label{sec.introduction}
Generative artificial intelligence (AI) is rapidly becoming part of higher education. Alongside widely available general-purpose chatbots, universities are beginning to provide institutionally managed AI platforms and course-specific tutoring systems at larger scales. These developments create new opportunities for immediate and individualized support, but they also raise a central educational question: Under which instructional conditions does conversational AI support learning rather than merely facilitate task completion? Addressing this question requires evaluations that are grounded in established principles of learning and instruction rather than in the technical capabilities of AI systems alone.

\textit{Worked examples} provide one such established instructional approach. By presenting a problem together with a structured solution, they can help learners identify relevant problem features, understand solution steps, and develop transferable problem-solving schemas \citep{SwellerCooper1985,AtkinsonEtAl2000,Renkl2014}. Conventional static worked examples, however, provide the same guidance to all learners and cannot respond to individual questions or difficulties. Conversational AI may extend this format by allowing learners to request explanations, clarify individual steps, and access additional support when needed. An AI-supported worked example may thus combine the structured guidance of example-based learning with the responsiveness of a chatbot.

Whether such conversational functionality provides additional instructional value depends critically on the comparison condition. A comparison with conventional course material, such as a lecture script, captures differences in instructional design as well as differences in AI use. In contrast, comparison with a matched static worked example that shares the same core solution structure allows the added value of conversational support to be examined more directly. This distinction is important because findings on educational chatbots and AI-supported tutoring remain heterogeneous and appear to depend strongly on how systems are designed and embedded in instruction \citep{MaEtAl2014,WollnyEtAl2021,KuhailEtAl2023,WuYu2024}. Evaluating AI-supported learning environments therefore requires both an appropriate non-AI comparison and outcomes that extend beyond performance, including learners’ cognitive and motivational experiences.

A further challenge concerns whether students use AI-supported learning materials when their use is voluntary. The availability of an AI-based learning environment does not necessarily result in its adoption. Students may value immediate and individualized assistance while simultaneously expressing concerns about unreliable information, overreliance on AI, academic integrity, or the loss of independent problem-solving skills \citep{JohnstonEtAl2024,BalabdaouiEtAl2024}. Existing research has often focused on general attitudes or intended use rather than on students’ actual choices between AI-supported and conventional learning materials in authentic course settings. Investigating both instructional effectiveness and voluntary material choice is therefore important for understanding the practical educational potential of AI-supported learning environments.

The present study addresses these issues in an undergraduate electrodynamics course, where students must coordinate conceptual, spatial, and mathematical reasoning when solving physics problems. In a controlled intervention on Gauss’s law, students learned with an AI-supported worked example (hereafter referred to as the intelligent worked example), a static worked example, or the lecture script. The conditions were compared with respect to concept-test performance, self-efficacy, cognitive load, usability, and perceived educational value. In a subsequent exploratory phase on Ampère’s law, students could freely choose between the available learning materials, and their choices were examined in relation to concerns and perceived barriers regarding AI use. By combining a controlled comparison with an investigation of voluntary uptake, the study examines both the added instructional value of an intelligent worked example and whether students choose to use it under more authentic learning conditions.

\section{Theoretical background}\label{sec.theorie}

\subsection{Worked examples in physics education}

Problem solving is a central component of university physics learning. For novices, however, solving unfamiliar problems independently can impose substantial demands because they have not yet developed the knowledge structures needed to guide an efficient solution. A \emph{worked example} presents a problem together with a structured solution, allowing learners to study how a solution is constructed rather than having to discover it through search. In well-structured domains such as mathematics and physics, worked examples can therefore support the initial acquisition of problem-solving skills more effectively than conventional problem solving, a finding known as the \emph{worked-example effect} \citep{SwellerCooper1985,AtkinsonEtAl2000,Renkl2014}.

Cognitive load theory provides a common explanation for this effect \citep{SwellerEtAl1998}. Because working memory is limited, novices may devote substantial resources to searching for a solution procedure. Worked examples reduce such unproductive search and can direct attention toward relations between problem features, underlying principles, and solution steps. Their instructional value therefore lies not merely in providing an answer, but in making the structure and reasoning of a solution accessible for learning.

Effective example study nevertheless requires learners to process this reasoning actively. Successful learners tend to generate \emph{self-explanations} that connect solution steps to underlying principles and anticipate how the solution will proceed \citep{chi1989self,Renkl1997}. In physics, productive self-explanations similarly involve identifying the relevant principle, explaining why it applies, and clarifying the purpose of mathematical operations \citep{GjerdeEtAl2022}. Simply adding explanations or prompts, however, does not reliably improve learning; their effectiveness depends on the cognitive activity they elicit \citep{WittwerRenkl2010,BarbieriEtAl2023}.

The amount of guidance required also changes with learners' developing expertise. Detailed worked solutions are particularly useful while problem-solving schemas are still weak, whereas the same information may become redundant as competence increases. Guidance can therefore be gradually withdrawn through completion tasks or faded worked examples \citep{RenklAtkinson2003}. This reflects the broader expertise-reversal effect, according to which instructional support that benefits novices may become unnecessary or burdensome for more knowledgeable learners \citep{ScheigerEtAl2023}.

These principles are particularly relevant in physics, where problem solving requires learners to coordinate conceptual understanding, representations, and mathematical procedures. Physics education research suggests that worked examples are most useful when they make the underlying problem structure and reasoning explicit rather than merely presenting calculations \citep{BadeauEtAl2017,SchmidtWeigandEtAl2009}. Thus, effective example-based learning depends both on promoting principle-based processing and on providing an appropriate level of guidance.

\subsection{Supporting worked example-based learning through chatbots}

A limitation of conventional static worked examples is that their explanations and prompts are fixed in advance. Large language models (LLMs) may extend this format by responding to learners' questions, requesting explanations, and providing additional clarification when needed. Within the ICAP framework, however, deeper engagement depends on what learners actually do rather than on the interface itself: conversational support may elicit constructive or interactive processing, but does not guarantee it \citep{ChiWylie2014}.

Recent studies suggest that responsive AI support can be instructionally useful, but that its benefits depend on how the support is structured. In physics education, \citet{BeckerEtAl2025} found that both an AI chatbot and a structured tiered-hint system reduced cognitive load more than static textbook-style explanations, while the chatbot showed no clear advantage over structured support. This illustrates why AI-supported environments should be evaluated against meaningful instructional comparison conditions rather than against conventional materials alone.

Conversational support also creates risks that are directly relevant to learning from worked examples. Learners may accept AI-generated explanations without sufficiently processing or verifying them. \citet{KruppEtAl2023}, for example, found evidence of uncritical reliance on ChatGPT-generated solutions. The ISAR model describes the possibility that AI intended to support learning instead reduces productive cognitive processing as an \emph{inversion} effect \citep{BauerEtAl2025}. In addition, fluent but scientifically incorrect explanations may be difficult for learners to identify \citep{DahlkemperEtAl2023}. Educational use of LLMs therefore requires a pedagogical strategy that promotes critical engagement rather than replacing learners' own reasoning \citep{KasneciEtAl2023}. A worked example used as the basis for an AI tutor must therefore be designed to elicit thinking rather than to substitute for it.

\subsection{Domain-specific context}

Gauss’s and Ampère’s laws provide compact relations between electromagnetic fields and their sources and can greatly simplify field calculations when appropriate symmetries are present. Their application, however, requires students to recognize relevant symmetries, distinguish fields from flux or circulation, identify enclosed sources, and select an appropriate integration surface or path. Research in physics education has documented persistent difficulties with these conceptual and representational demands \citep{PepperEtAl2012,CamposEtAl2023,HernandezEtAl2023}. These characteristics make Gauss’s and Ampère’s laws suitable contexts for investigating structured problem-solving support.

\subsection{The present study and research questions}\label{sec}

The present study compares an intelligent worked example with a corresponding static worked example and the regular lecture script. Both worked examples use the same underlying solution structure and activation questions intended to promote self-explanation. In the intelligent worked example, an LLM presents the instructor-designed solution conversationally and provides additional support by responding to students’ questions, offering further explanations, and discussing their responses to the activation questions. The comparison with the lecture script therefore examines the overall instructional value of the AI-supported worked-example format, whereas the comparison with the static worked example more specifically examines the added value of conversational support while keeping the core instructional structure constant.

In a subsequent exploratory phase, students could freely choose among the available learning materials. This complements the controlled comparison with an examination of whether AI-supported materials are voluntarily adopted and whether material choice is related to concerns and perceived barriers regarding AI use.

\begin{itemize}
\item []\textbf{RQ1 (Group differences):}\
How do students who learn with an intelligent worked example, a static worked example, or the lecture script differ with respect to
\begin{itemize}[leftmargin=2.5em]
\item[(a)] concept-test performance and learning gains immediately after the intervention and at follow-up?
\item[(b)] changes in self-efficacy?
\item[(c)] perceived cognitive load during learning?
\item[(d)] perceived usability and educational value of the learning environment?
\end{itemize}
\end{itemize}

\begin{itemize}
\item []\textbf{RQ2 (Material choice and AI-related concerns):}\
Which learning materials do students choose when their use is voluntary, and how is this choice related to concerns and perceived barriers regarding AI use?
\end{itemize}

\section{Methods}\label{sec.methods}

The following sections describe the study design and learning materials (Sect.~\ref{sec:study}), the sample (Sect.~\ref{sec:sample}, Tab.~\ref{Tab.sample}), and the statistical methods used to address the research questions (Sect.~\ref{sec:analysis}).

\subsection{Study and materials}\label{sec:study}

The overall study procedure and measurement occasions are illustrated in Fig.~\ref{Fig.: Procedure}. An overview of the instruments and data collected at each measurement occasion is provided in Tab.~\ref{tab: Instrumente}. A detailed description of the instruments is provided in Sect.~\ref{sect.measures}.

\subsubsection{Study design}

\begin{figure}[htbp]
  \begin{center}
    \includegraphics[width=\textwidth]{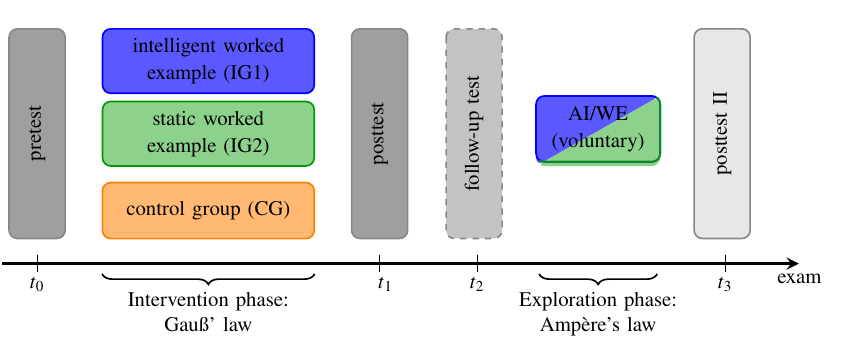}
  \end{center}
  \caption{Overview of the study design and measurement occasions from pretest ($t_0$) to the assessment following the exploration phase ($t_3$; IG1 = intelligent worked example group, IG2 = static worked example group, CG = control group)}
  \label{Fig.: Procedure}
\end{figure}

The study was conducted in the undergraduate Experimental Physics II course at the University of Göttingen, which typically introduces physics students to university-level electromagnetism in their second semester. The course combines lectures with weekly recitation sessions in which compulsory assignments are discussed.

The study comprised two consecutive phases: a controlled intervention on Gauss’s law and a subsequent exploratory phase on Ampère’s law. Participants provided written informed consent when completing a study questionnaire or test for the first time. Self-generated pseudonymous codes were used to link responses across measurement occasions.

\paragraph{Intervention phase -- Gauss's law}
The intervention followed a cluster-randomized pretest--posttest--follow-up design with three learning conditions and was conducted during the regular recitation sessions. Students had selected fixed recitation groups before the learning conditions were assigned. The nine groups were subsequently randomized to the intelligent worked-example condition (IG1), the static worked-example condition (IG2), or the lecture-script condition (CG), with three groups assigned to each condition while accounting for group-size differences.

Each session lasted 90 minutes. Students first completed a pretest ($t_0$) assessing conceptual understanding of Gauss’s law and self-efficacy regarding its application. Gauss’s law had previously been introduced in the lecture. Students then worked individually and at their own pace on the same target problem, supported by the material assigned to their condition. Students in both worked-example conditions could additionally consult the lecture script. Material use was self-directed, and no additional instructional support was provided during the learning phase.

Immediately afterward, students completed the posttest ($t_1$), including the same concept test and self-efficacy measure as at pretest, as well as measures of cognitive load, usability, and perceived educational value.

Four weeks later, the concept test was administered again during the regular lecture ($t_2$). Because participation at follow-up was independent of participation in the intervention, some students completed only this assessment. These students constituted a non-equivalent cross-sectional reference group and were not included in longitudinal analyses of learning gains.

\paragraph{Exploration phase -- Ampère's law}
The subsequent exploration phase focused on Ampère's law. Students completed a regular homework assignment requiring them to apply Ampère's law and could freely choose whether to use an intelligent worked example, a static worked example, both worked examples, or neither as additional learning support. The regular lecture script remained available throughout the course and was not provided specifically for this assignment. Unlike the intervention phase, the target problem was completed outside the supervised recitation setting and submitted as part of the regular course work.

After submitting the assignment, students reported which of the two worked examples they had used and completed questionnaires assessing their concerns about AI use and perceived barriers to using AI-based learning support ($t_3$; Fig.~\ref{Fig.: Procedure}). The exploration phase followed an observational design: students freely selected their learning materials, and neither the timing nor the intensity of material use was controlled or monitored. Information on the use of the worked examples was therefore based on students' self-reports.

\subsubsection{Measures and materials}\label{sect.measures}
This section describes the learning materials used during the intervention and exploration phases, as well as the instruments employed to assess students' conceptual understanding, self-efficacy, cognitive load, usability and perceived educational value, and AI-related concerns and barriers. All learning materials and assessment instruments are provided in Appendices~\ref{secA} and~\ref{secA1}. An overview of all scales is given in Tab.~\ref{tab: Instrumente}.

\begin{figure}[htbp]
  \centering

  \begin{tikzpicture}
    \node[anchor=south west, inner sep=0] (img) at (0,0)
      {\includegraphics[width=\textwidth]{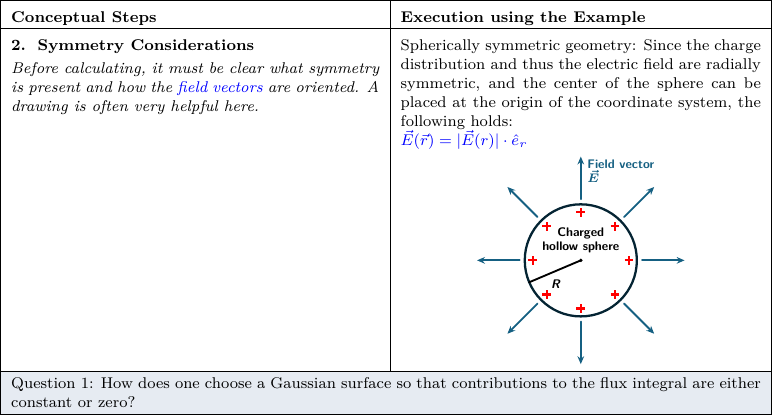}};
    \node[anchor=south west, font=\bfseries]
      at ([xshift=0mm,yshift=0mm]img.north west) {(a)};
  \end{tikzpicture}
  \begin{tikzpicture}
    \node[anchor=south west, inner sep=0] (img) at (0,0)
      {\includegraphics[width=\textwidth]{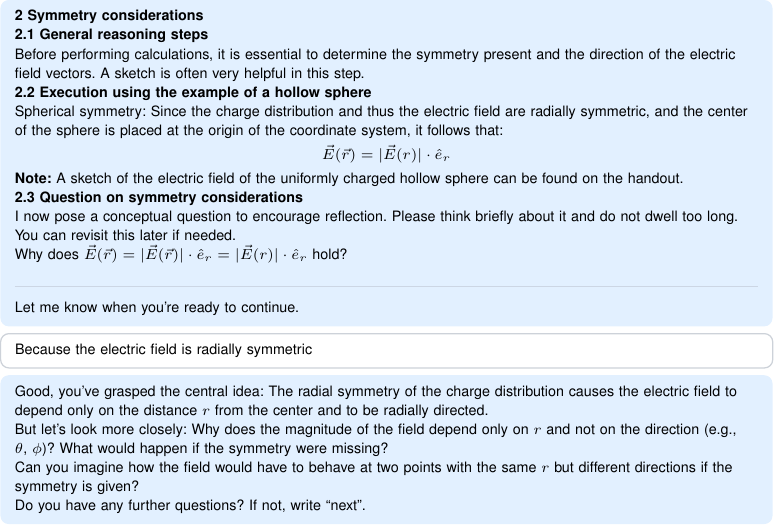}};
    \node[anchor=south west, font=\bfseries]
      at ([xshift=0mm,yshift=0mm]img.north west) {(b)};
  \end{tikzpicture}\vspace{-3mm}
\caption{Static worked example (a) and intelligent worked example (b). In the static worked example, the activation question serves as a prompt for self-explanation, without providing feedback or further interaction. In the intelligent worked example, the same type of question can initiate a discussion with the chatbot. More generally, the intelligent worked example allows students to ask follow-up questions and receive additional support during learning. The complete static worked example is provided in Appendix~\ref{secWE}. The intelligent worked example is reproduced in English for illustrative purposes; the original version used in the study was in German, and the corresponding system prompt is provided in Appendix~\ref{secPrompt}}
  \label{Fig.: WE}
\end{figure}

\paragraph{Worked examples}
Students in both intervention groups learned to apply Gauss's law using worked examples developed specifically for this study (Fig.~\ref{Fig.: WE} and Appendix~\ref{secWE}). Both versions addressed the same learning objectives and presented an identical step-by-step solution for calculating the electric field of a hollow sphere using Gauss's law. The worked examples were designed according to established principles of example-based learning \citep{AtkinsonEtAl2000,Renkl2014}. The solution was structured into clearly separated subgoals, each accompanied by a brief conceptual explanation. Activation questions were integrated at selected points in the solution process to prompt students to reflect on the underlying physical concepts and engage in self-explanation \citep{chi1989self}. The questions were open-ended and required students to formulate an explanation in their own words.

The static worked example (Fig.~\ref{Fig.: WE}a) was provided digitally as a PDF and presented the complete worked solution in a tabular format. For the complete static worked example see Appendix~\ref{secWE}. The individual solution steps and corresponding conceptual explanations were arranged sequentially, with the activation questions embedded directly at the relevant points in the table. Students worked through the material independently and answered these questions without receiving feedback or additional support.

The intelligent worked example (Fig.~\ref{Fig.: WE}b) extended the static version by integrating an AI-based chatbot into the learning environment. The worked example followed the same sequence of solution steps and activation questions, but provided conversational support throughout the learning process. Students could use the chatbot to ask follow-up questions, request additional explanations, and discuss their responses to the activation questions before proceeding to the next solution step. Thus, the intelligent and static worked examples were identical in terms of content, solution structure, and instructional explanations, differing only in the availability of conversational, AI-based support.

The intelligent worked example was implemented using ChatAI, the institutional AI platform provided by the University of Göttingen \citep{UniGoettingen2026ChatAI}. The underlying large language model was Qwen3-30B-A3B-Instruct-2507, with a temperature of 0.0 and a top-p value of 0.05. The intervention was conducted between May 5 and 7, 2026. The system prompt used to guide the chatbot’s responses is provided in Appendix~\ref{secPrompt}.

During the exploration phase, students had access to both an intelligent and a static worked example on calculating the magnetic field of a long solenoid using Ampère's law. Both were developed according to the same instructional design principles and followed the same structure as their respective counterparts on Gauss's law used during the intervention phase.

\begin{figure}[htbp]
  \begin{center}
    \fbox{\includegraphics[width=0.7\textwidth]{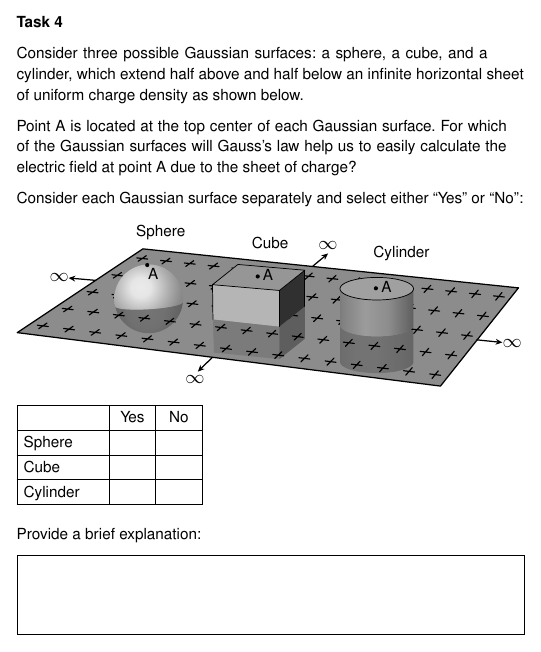}}
  \end{center}
  \caption{Example task adapted from the concept test by \citet{singh2006student}, comprising three true--false statements and one explanatory item. A list of all items in the concept test is provided in Appendix~\ref{secA1.2}}
  \label{Fig.: Test2}
\end{figure}

\paragraph{Concept test on Gauss's law}
Students' conceptual understanding of Gauss's law was assessed using a paper-and-pencil concept test. The test comprised five tasks with several subtasks, yielding 19 items in total (Appendix~\ref{secA1.2}). The tasks covered key concepts related to Gauss's law, including the identification of suitable Gaussian surfaces, and were adapted from an established concept test on symmetry and Gauss's law developed by \citet{singh2006student}. The original single-best-answer multiple-choice items were transformed into a multiple true--false format by treating each response option as an individual statement to be judged as true or false, resulting in 18 true--false items. In addition, one open-ended item required students to justify their reasoning. An example task is shown in Fig.~\ref{Fig.: Test2}.

To combine the two test components into an overall score, the composite score was defined as $T(\kappa)=T_1+\kappa T_2$, where $T_1$ denotes the score on the 18 true--false statements, $T_2$ the score on the explanatory item, and $\kappa$ determines the relative weight of the explanatory item. The weighting parameter was specified a priori as $\kappa=2$, such that each point awarded for the explanatory item contributed twice the weight of a correctly answered true--false statement.

\begin{figure}[htbp]
  \begin{center}
    \includegraphics[width=0.8\textwidth]{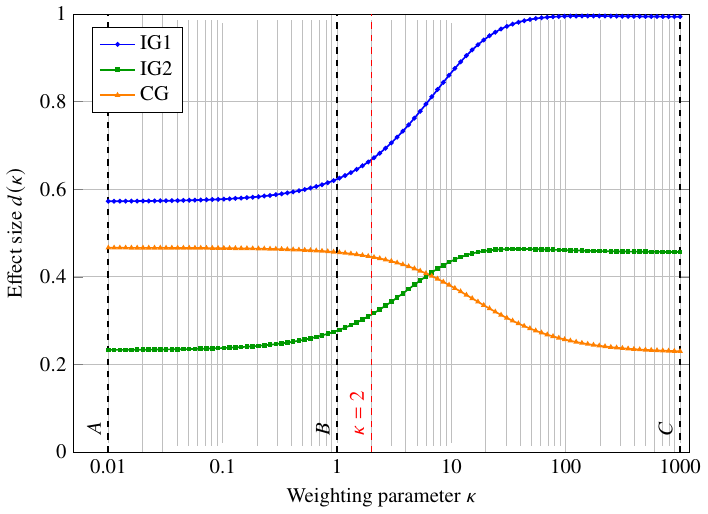}
  \end{center}
  \caption{Test combination analysis showing the pre--post effect size (Cohen's $d$) of the composite test score $T=T_1+\kappa T_2$ for the intelligent worked example group (IG1), the static worked example group (IG2), and the control group (CG) as a function of the weighting parameter $\kappa$. Here, $T_1$ denotes the score on the true--false items and $T_2$ the score on the explanatory item. For each student, composite scores were calculated from the raw pre- and posttest scores for 100 different values of $\kappa$. The $\kappa$ axis is logarithmic to provide sufficient resolution across the relevant range of weighting values. The vertical dashed lines indicate three reference cases: $A$, the composite score is dominated by the true--false items ($\kappa\rightarrow0$); $B$, scores from both test components receive equal unit weight ($\kappa=1$); and $C$, the composite score is dominated by the explanatory item ($\kappa\rightarrow\infty$). The red dashed line marks the prespecified weighting parameter $\kappa=2$, which was used to calculate response accuracy in all subsequent analyses}
  \label{Fig.: Test}
\end{figure}

To examine the robustness of this prespecified weighting, a sensitivity analysis was conducted following \citet{klein2017}. Pre--post effect sizes (Cohen's $d$) were calculated separately for the three groups across a range of values of $\kappa$. The analysis was not used to select the weighting parameter, but to examine where the prespecified value $\kappa=2$ was located within the range of alternative weighting regimes and how sensitive the resulting group-specific effect sizes were to changes in $\kappa$. Cohen's $d$ was used specifically for this sensitivity analysis following \citet{klein2017}, whereas generalized eta squared ($\eta_G^2$) was used for the ANOVA-based analyses reported in Sect.~\ref{sec:analysis}. The results of the sensitivity analysis are shown in Fig.~\ref{Fig.: Test}.

For all subsequent analyses, response accuracy was calculated using the prespecified weighting $\kappa=2$:
\begin{equation*}
\text{response accuracy}=\frac{X+2p}{24},
\end{equation*}
where $X \in {0,\dots,18}$ denotes the number of correctly answered true--false statements and $p \in {0,\dots,3}$ the score on the explanatory item. The explanatory item therefore contributed a maximum of six points to the total score of 24, and response accuracy ranged from $0$ to $1$.

\paragraph{Self-efficacy questionnaire}
Students' self-efficacy regarding the application of Gauss's law was assessed using a four-item questionnaire (Appendix~\ref{secA1.3}). The items were developed following the guidelines for constructing self-efficacy scales proposed by \citet{bandura2006guide} and informed by the instrument developed by \citet{freed2022development}. They addressed students' perceived ability to understand the concept of Gauss's law, identify suitable charge symmetries, construct appropriate Gaussian surfaces, and apply Gauss's law to calculate electric fields. An example item is \textit{``I can assess whether the symmetry of the charge distribution is suitable for applying Gauss's law.''} Responses were recorded on a six-point Likert-type scale ranging from 1 (\textit{strongly disagree}) to 6 (\textit{strongly agree}).

\paragraph{Cognitive load questionnaire}
Students' perceived cognitive load during the self-learning phase was assessed using the validated cognitive load questionnaire developed by \citet{leppink2013}. The original instrument distinguishes three dimensions of cognitive load---intrinsic cognitive load (ICL), extraneous cognitive load (ECL), and germane cognitive load (GCL)---using three or four items per subscale. Its three-factor structure has been shown to be robust across adaptations to different disciplinary contexts \citep{leppink2014}.

For the present study, the questionnaire was adapted to the context of learning Gauss's law and comprised nine items (Appendix~\ref{secA1.3}). Example items were \textit{``The topic `Calculating an electric field using Gauss's law' was very complex.''} (ICL), \textit{``The instructions and/or explanations during the self-learning phase were, in terms of learning, very ineffective.''} (ECL), and \textit{``The self-learning phase really enhanced my understanding of the topic `Calculating an electric field using Gauss's law'.''} (GCL). Responses were recorded on a six-point Likert-type scale ranging from 1 (\textit{strongly disagree}) to 6 (\textit{strongly agree}).

\paragraph{Questionnaire on usability and perceived educational value}
Usability and perceived educational value of the learning materials were assessed using two three-item scales (Appendix~\ref{secA1.3}). The usability scale focused on the clarity, comprehensibility, and ease of use of the learning materials. Its items were adapted from the System Usability Scale \citep[SUS;][]{brooke1996sus} and the validated adaptation by \citet{unver2017reliability}. An example item is \textit{``The representation format of the learning material is clear and understandable.''}

Perceived educational value was assessed using three items adapted from \citet{shellman2006simulations}. The items addressed students' perceptions of the usefulness of the learning materials for learning and their potential use in other physics courses. An example item is \textit{``I recommend using the learning material in other courses (e.g., Experimental Physics I) and extending it to further concepts covered in Experimental Physics II.''} Responses to both scales were recorded on a six-point Likert-type scale ranging from 1 (\textit{strongly disagree}) to 6 (\textit{strongly agree}).

\paragraph{Choice of learning materials}
During the exploration phase, students freely chose which of the two learning materials to use. Based on their self-reported use, they were categorized as having used the intelligent worked example, the static worked example, both worked examples, or neither. As the regular course script remained available throughout the course, its use was not assessed separately.

\paragraph{Questionnaire on AI-related concerns and barriers}
Students' concerns about using AI were assessed using five items developed for the present study: two with reference to the perceived risk of use subscale of the TAME-ChatGPT instrument \citep{sallam2024chatgpt}, one to a study of ChatGPT use for homework assignments \citep{miljkovic2024}, one to examples of flawed AI-generated physics content \citep{bitzenbauer2023}, and one without reference to a published source. All items continued the stem \textit{``I am concerned that \ldots''} and addressed potential effects of AI use on independent problem solving and critical thinking, as well as concerns about unnoticed errors and superficial understanding when learning physics with AI. An example item is \textit{``I am concerned that using AI may reduce my ability to solve problems independently.''}

Perceived barriers to AI use were assessed using five items developed for the present study and conceptually related to perceived ease of use \citep{davis1989}. The items addressed difficulties in identifying relevant information, text-heavy responses, prompt formulation, the perceived ratio of effort to benefit, and the user interface. An example item is \textit{``The AI's responses often contain too much text and are therefore difficult to navigate.''}

Responses to both scales were recorded on a seven-point Likert-type scale ranging from 1 (\textit{strongly disagree}) to 7 (\textit{strongly agree}). Neither scale referred specifically to the intervention or its learning materials; they were therefore intended to capture students' general concerns about and perceived barriers to AI use rather than evaluations of the specific learning environment (Appendix~\ref{secA1.4}).

\subsection{Sample} \label{sec:sample}

\begin{table}[htbp]
\caption{Sample data (intervention groups IG1 (intelligent worked example) and IG2 (static worked example), control group CG, number No., statistical significance $p$ using one-way ANOVA). The total sample comprises students who completed the intervention phase, pre-, and posttest.}
\label{Tab.sample}
\centering
  \renewcommand*{\arraystretch}{1.3}
	\begin{tabular}{p{3.4cm}ccccc}	
     \hline \hline 
  &Total& IG1  & IG2 & CG  & $p$ \\\hline  
No. of subjects &  $86^{\text{a,b}}$ &$29$ & $28$ &$29$& -- \\ 
No. of female subjects & $31$ ($36\%$) &$6$ ($21\%$) &$11$ ($39\%$)&$14$ ($48\%$) & --\\ 
Mean age (in years) & $20.4 \pm 2.9$ &$20.4 \pm 2.3$& $20.3 \pm 2.8$ &$20.6 \pm 3.5$& $0.89$ \\ 
No. of semesters studied & $2.4 \pm 1.7$ &$3.1 \pm 2.8$& $2.0 \pm 0.2$&$2.0 \pm 0.4$& $0.15^{\text{c}}$\\
Average grade for university entrance$^{\text{d}}$ & $1.6 \pm 0.6$ &$1.8 \pm 0.8$&$1.4 \pm 0.5$&$1.7 \pm 0.5$ & $0.08^{\text{c}}$\\ 
Pretest score Gauss's law$^{\text{e}}$ & $0.61 \pm 0.19$ &$0.58 \pm 0.21$&$0.66 \pm 0.15$&$0.59 \pm 0.20$ & $0.19$\\ 
\hline
\multicolumn{6}{p{\textwidth}}{$^{\text{a}}$A subsample of $N = 57$ comprised students who completed all three test points ($N_{\text{IG1}} = 21$, $N_{\text{IG2}} = 20$, $N_{\text{CG}} = 16$). An additional $N = 13$ students completed only the follow-up test.}\\
\multicolumn{6}{p{\textwidth}}{$^{\text{b}}$A subsample of $N = 46$ comprised students who also participated in the exploration phase ($N_{\text{IG1}} = 16$, $N_{\text{IG2}} = 13$, $N_{\text{CG}} = 17$). An additional $N = 6$ students participated only in the exploration phase.}\\
\multicolumn{6}{l}{$^{\text{c}}$Welch's ANOVA due to violation of homogeneity of variances was performed.}\\
\multicolumn{6}{l}{$^{\text{d}}$The scale ranges from $1.0$ (best performance) to $4.0$. The grades are indicated by the students.}\\
\multicolumn{6}{l}{$^{\text{e}}$The scale ranges from $0$ to $1$ (best performance).}\\
 \hline \hline 

\end{tabular}
\end{table}

The analyses of the intervention phase (RQ1) were based on a core sample of $86$ students who completed the intervention as well as the pretest and immediate posttest. This sample comprised $29$ students in the intelligent worked-example group (IG1), $28$ in the static worked-example group (IG2), and $29$ in the control group (CG; see Tab.~\ref{Tab.sample}). No statistically significant baseline differences were observed between the three groups in age, number of semesters studied, university entrance grade, or pretest performance.

For the longitudinal follow-up analyses of RQ1(a)(ii), complete data across pretest, posttest, and follow-up test were available for $57$ students ($n_{\mathrm{IG1}}=21$, $n_{\mathrm{IG2}}=20$, $n_{\mathrm{CG}}=16$). An additional $13$ students who had not participated in the intervention completed the follow-up assessment. Because these students had neither been randomly assigned to a study condition nor completed the pretest and immediate posttest, they were included only as a non-equivalent cross-sectional reference group at follow-up and were excluded from analyses of individual learning gains. Thus, longitudinal analyses were based on $N=57$, whereas exploratory cross-sectional comparisons at follow-up included $N=70$ students.

For RQ2, $46$ students from the intervention sample also participated in the subsequent exploration phase ($n_{\mathrm{IG1}}=16$, $n_{\mathrm{IG2}}=13$, $n_{\mathrm{CG}}=17$). A further $6$ students participated only in the exploration phase, resulting in a total sample of $N=52$ for analyses of material choice. Because of missing questionnaire responses, the available samples for analyses of AI-related concerns and perceived barriers were $N=41$ and $N=40$, respectively.

\subsection{Data analysis} \label{sec:analysis}
All variables were linearly transformed to a common range of $[0,1]$. Distributions of the continuous measures were considered sufficiently close to normal for parametric analyses based on their skewness and kurtosis and the sample size. Internal consistencies of the multi-item scales are reported in Tab.~\ref{tab: Instrumente}; results involving the extraneous cognitive load scale, which showed comparatively low reliability ($\alpha_C=.56$), are interpreted with caution.

For RQ1, between-group differences at individual measurement points were examined using one-way ANOVAs, and changes over time using repeated-measures and mixed-design ANOVAs. The mixed-design analyses tested time $\times$ group effects for response accuracy and self-efficacy, load type $\times$ group effects for cognitive load, and scale $\times$ group effects for usability and perceived educational value. Where appropriate, Tukey-adjusted pairwise comparisons were conducted. Generalized eta squared ($\eta_G^2$) was reported for ANOVA-based analyses \citep{olejnik2003generalized}, and Cohen's $d$ for significant pairwise differences.

Individual change scores were additionally calculated for response accuracy and self-efficacy from pretest to posttest and, for response accuracy in the longitudinal follow-up sample, from posttest to follow-up. Group differences in change scores were examined using one-way ANOVAs. Learning gains in response accuracy from pretest to posttest were also described using Hake's normalized gain $g_H$ \citep{Hake1998}.

For RQ2, material-choice groups were small and substantially unequal in size. Differences in AI-related concerns and perceived barriers among the four choice groups were therefore examined using Kruskal--Wallis tests. Students who used the intelligent worked example, either alone or together with the static worked example, were additionally compared with students who did not use it using Mann--Whitney $U$ tests.

Pearson correlations were used to examine associations between continuous measures. Statistical tests were evaluated at $\alpha=.05$. Pairwise comparisons were adjusted as described above.

\begin{table}[htb]
\caption{Overview of scales included for data analysis (Cronbach's alpha $\alpha_C$ based on the respective sample).}
    \label{tab: Instrumente}
    \centering
    \renewcommand*{\arraystretch}{1.3}
    \begin{tabular}{p{3cm}p{6.8cm}cc}
    \hline \hline 
   Time point & Scale & \#Items & $\alpha_C$\\ 
   \hline
   $t_0$ (pretest) & Concept test (response accuracy)$^{\text{a}}$ & $19$ &  $0.78$  \\
 & Self-efficacy  & 4 & $0.87$ \\
& Sociodemographics & -- & -- \\
 $t_1$ (posttest) 
 & Concept test (response accuracy)$^{\text{a}}$  & $19$ &  $0.78$\\
& Self-efficacy & 4 & $0.90$ \\
& Intrinsic cognitive load (ICL) & 3 & $0.77$ \\
& Extraneous cognitive load (ECL) & 3 & $0.56$ \\
& Germane cognitive load (GCL) & 3 & $0.89$ \\
& Usability  & 3 & $0.68$ \\
& Perceived educational value  & 3 & $0.78$ \\

 $t_2$ (Follow-up test) & Concept test (response accuracy)$^{\text{a}}$  & $19$ &  $0.74$\\
  $t_3$ (posttest II) & Choice of learning material$^{\text{b}}$ & -- & -- \\
& Concerns towards AI  & 5 & $0.62$ \\
& Barriers to AI use  & 5 & $0.77$ \\
\hline
 \multicolumn{4}{p{\textwidth}}{$^{\text{a}}$Eighteen true--false statements and one open-ended explanatory item; $\alpha_C$ was computed over these 19 items. In the test score the explanatory item carries the weight $\kappa = 2$ (Section~\ref{sect.measures}); this weighting does not enter the reliability reported here.}\\
 \multicolumn{4}{p{\textwidth}}{$^{\text{b}}$Self-reported after the exploration phase; not a scale, so no item count and no reliability are reported.}\\

                          \hline\hline
             
    \end{tabular}
\end{table}

\section{Results}\label{sec.results} 

\subsection{RQ1 -- Group differences}

\subsubsection{Performance (a)}

\begin{figure}[htbp]
  \begin{center}
        \includegraphics[width=0.8\textwidth]{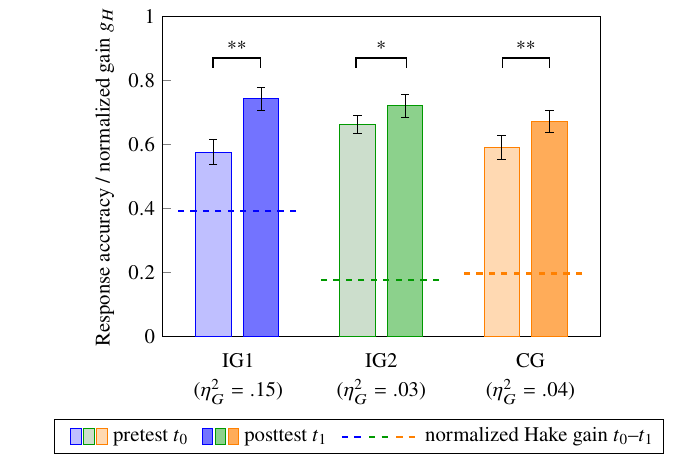}
    \end{center}
    \caption{Comparison of students' response accuracy and learning gain for the intelligent worked example group (IG1), the static worked example group (IG2), and the control group (CG) from pre- ($t_0$) to posttest ($t_1$; $N=86$). Normalized gain $g_H$ from pre- to posttest according to \citet{Hake1998} for all groups is visualized with dashed lines. Response accuracy between $t_0$ and $t_1$ is compared using one-way repeated measures ANOVA ($^{\ast}$~/~$^{\ast\ast}$ statistical significance $p<.05$~/$p<.01$, effect size $\eta_G^2$, error bars represent 1 SEM)}\label{Fig.: Performancea}
\end{figure}

From pretest ($t_0$) to posttest ($t_1$), the time $\times$ group interaction was not significant ($p=.07$), indicating no significant differences in response accuracy changes among the intelligent worked example group (IG1), the static worked example group (IG2), and the control group (CG). Across all three groups, response accuracy increased significantly over time [$F(1,83)=26.24$, $p<.001$, $\eta_G^2=.07$; Fig.~\ref{Fig.: Performancea}]. In IG1, response accuracy increased from $0.58\pm0.21$ to $0.74\pm0.19$, corresponding to a large effect [$F(1,28)=12.61$, $p=.001$, $\eta_G^2=.15$]. IG2 showed a smaller but significant increase from $0.66\pm0.15$ to $0.72\pm0.19$ [$F(1,27)=6.74$, $p=.015$, $\eta_G^2=.03$], corresponding to a small effect. Similarly, response accuracy in the control group increased from $0.59\pm0.20$ to $0.67\pm0.18$, also corresponding to a small effect [$F(1,28)=8.01$, $p=.009$, $\eta_G^2=.04$]. Tukey-adjusted pairwise comparisons of change scores revealed no significant differences between the groups.

Consistent with this descriptive pattern, the normalized gain according to \citet{Hake1998} was medium in IG1 ($g_{H,\mathrm{IG1}}=.39$) and small in both IG2 ($g_{H,\mathrm{IG2}}=.18$) and the control group ($g_{H,\mathrm{CG}}=.20$; Fig.~\ref{Fig.: Performancea}, dashed). Despite these differences in descriptive learning gains, posttest response accuracy did not differ significantly between the three groups ($p=.34$).

The test combination analysis (Sect.~\ref{sect.measures}) further showed that pre--post effect sizes in both worked example groups increased with greater weighting of the explanatory item, whereas the opposite pattern emerged in the control group (Fig.~\ref{Fig.: Test}). Thus, learning gains in IG1 and IG2 were more pronounced for the explanatory component, while those in the control group were more pronounced for the true--false component.

\begin{figure}[htbp]
  \begin{center}
        \includegraphics[width=0.75\textwidth]{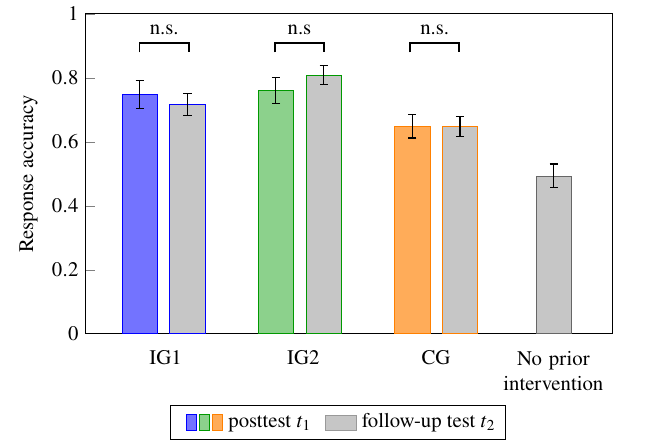}
    \end{center}
    \caption{Comparison of students' response accuracy ($N=70$ including 13 students who had not completed the posttest) for the intelligent worked example group (IG1), the static worked example group (IG2) and the control group (CG) in post- ($t_1$) and follow-up test ($t_2$). Follow-up test score of students who did not complete the intervention phase is visualized ("No prior intervention").  Response accuracy between the time points ($t_1$, $t_2$) is compared using one-way repeated measures ANOVA (n.s. not statistically significant, error bars represent 1 SEM)}\label{Fig.: Performanceb}
\end{figure}

When the follow-up test ($t_2$) was included, pretest--posttest scores and learning effects in the reduced sample ($N=57$) were comparable to those observed in the full sample. From posttest to follow-up, the time $\times$ group interaction was not significant ($p=.13$), indicating no significant differences in changes in response accuracy between the groups. However, a significant main effect of group was observed, corresponding to a medium-sized effect [$F(2,54)=3.69$, $p=.031$, $\eta_G^2=.12$; Fig.~\ref{Fig.: Performanceb}]. In IG1, response accuracy decreased slightly from $0.75\pm0.20$ at posttest to $0.72\pm0.16$ at follow-up, although this change was not significant ($p=.16$). In contrast, IG2 showed a further, albeit non-significant, increase from $0.76\pm0.18$ to $0.81\pm0.13$ ($p=.11$), while response accuracy in the control group remained stable at $0.65$ ($0.65\pm0.15$ at posttest and $0.65\pm0.13$ at follow-up; $p=1.00$). Tukey-adjusted pairwise comparisons of change scores revealed no significant differences between the groups.

While response accuracy did not differ significantly between the three groups at posttest ($p=.14$), a significant group difference emerged at follow-up, with a large effect [$F(2,54)=5.76$, $p=.005$, $\eta_G^2=.18$]. Tukey-adjusted pairwise comparisons revealed significantly higher follow-up scores in IG2 than in the control group, corresponding to a large effect ($p=.004$, $d=1.22$), whereas the remaining pairwise differences were not significant.

Students who did not complete the intervention phase ($N=13$) achieved a mean follow-up score of $0.49\pm0.13$. Including these students in the follow-up comparison resulted in an even larger overall group effect [$F(3,66)=13.79$, $p<.001$, $\eta_G^2=.39$]. Tukey-adjusted pairwise comparisons showed significantly higher follow-up scores for IG1 ($p<.001$, $d=1.48$), IG2 ($p<.001$, $d=2.33$), and the control group ($p=.023$, $d=1.18$) compared with students who did not complete the intervention phase, with all three differences corresponding to large effects. In addition, IG2 scored significantly higher than the control group ($p=.007$, $d=1.22$), also corresponding to a large effect, whereas the remaining pairwise differences were not significant.

\subsubsection{Self-efficacy (b)}

\begin{figure}[htbp]
  \begin{center}
        \includegraphics[width=0.6\textwidth]{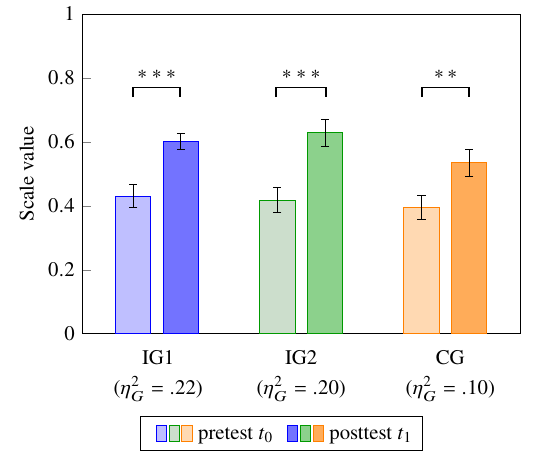}
    \end{center}
    \caption{Comparison of students' self-efficacy for the intelligent worked example group (IG1), the static worked example group (IG2), and the control group (CG) from pre- ($t_0$) to posttest ($t_1$; $N=86$). Self-efficacy between $t_0$ and $t_1$ is compared using one-way repeated measures ANOVA ($^{\ast\ast}$~/~$^{\ast\ast\ast}$ statistical significance $p<.01$~/$p<.001$, effect size $\eta_G^2$, error bars represent 1 SEM)}\label{Fig.: SE}
\end{figure}

For self-efficacy regarding concepts related to Gauss's law, the time $\times$ group interaction from pretest ($t_0$) to posttest ($t_1$) was not significant ($p=.38$), indicating no significant differences in changes in self-efficacy between the groups. Across all three groups, self-efficacy increased significantly over time [$F(1,83)=71.25$, $p<.001$, $\eta_G^2=.16$; Fig.~\ref{Fig.: SE}]. This increase corresponded to large effects in the intelligent worked example group [$F(1,28)=24.71$, $p<.001$, $\eta_G^2=.22$] and the static worked example group [$F(1,27)=53.13$, $p<.001$, $\eta_G^2=.20$], and to a medium-sized effect in the control group [$F(1,28)=11.06$, $p=.002$, $\eta_G^2=.10$]. Tukey-adjusted pairwise comparisons of change scores likewise revealed no significant differences between the groups. Self-efficacy did not differ significantly between the groups at either pretest ($p=.79$) or posttest ($p=.19$). Finally, a significant large positive correlation between posttest self-efficacy and posttest response accuracy was observed only in the control group ($r=.58$, $p=.001$).

\subsubsection{Perceived cognitive load (c)}

\begin{figure}[htbp]
  \begin{center}
        \includegraphics[width=0.9\textwidth]{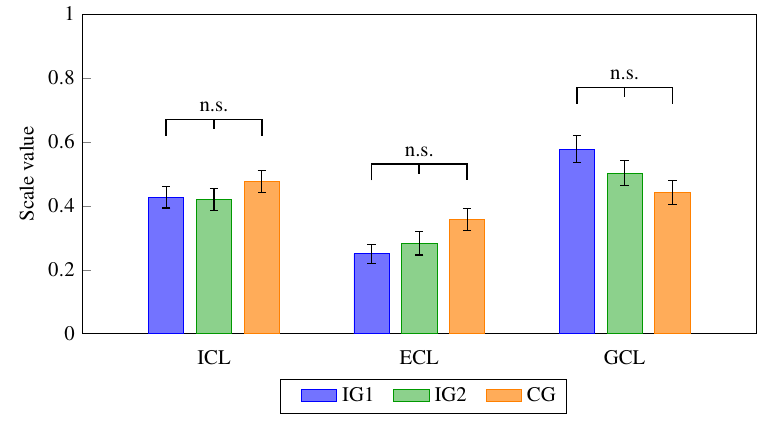}
    \end{center}
    \caption{Students' perceived cognitive load for the intelligent worked example group (IG1), the static worked example group (IG2), and the control group (CG; $N=86$). Extraneous cognitive load (ECL), intrinsic cognitive load (ICL), and germane cognitive load (GCL) are compared between the groups using one-way ANOVAs with group as the between-subject factor (n.s. not statistically significant, error bars represent 1 SEM)}\label{Fig.: CL}
\end{figure}

For perceived cognitive load, a significant cognitive load type $\times$ group interaction was observed, with cognitive load type (intrinsic, extraneous, and germane cognitive load) as the within-subjects factor and group as the between-subjects factor [$F(4,166)=3.15$, $p=.016$, $\eta_G^2=.05$]. This small interaction effect was mainly reflected in the descriptive patterns of extraneous cognitive load (ECL) and germane cognitive load (GCL): IG1 reported the lowest ECL and the highest GCL, whereas the control group showed the opposite pattern, with the highest ECL and the lowest GCL (Fig.~\ref{Fig.: CL}). When examining the cognitive load types separately, no significant overall group differences were found for intrinsic cognitive load ($p=.46$), ECL ($p=.08$), or GCL ($p=.06$). Tukey-adjusted pairwise comparisons likewise revealed no significant differences between the groups for intrinsic cognitive load or ECL. For GCL, however, IG1 reported significantly higher values than the control group, corresponding to a medium-sized effect ($p=.046$, $d=.63$).

\subsubsection{Usability and perceived educational value (d)}

\begin{figure}[htbp]
  \begin{center}
        \includegraphics[width=0.6\textwidth]{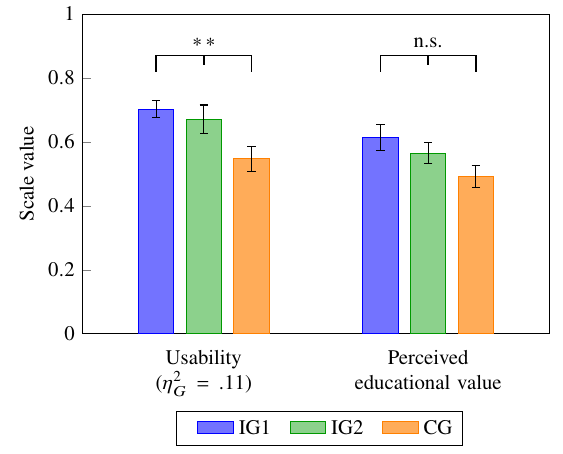}
    \end{center}
    \caption{Indicated usability and students' perceived educational value of the learning environment for the intelligent worked example group (IG1), the static worked example group (IG2), and the control group (CG; $N=86$). Usability and perceived educational value are compared between the groups using one-way ANOVAs with group as the between-subject factor ($^{\ast\ast}$~/~n.s. statistical significance $p <.01$~/~not significant, effect size $\eta_G^2$, error bars represent 1 SEM)}\label{Fig.: Usability}
\end{figure}

For usability and perceived educational value, the scale $\times$ group interaction was not significant ($p=.67$). Across both scales, however, a significant group effect was observed, corresponding to a medium-sized effect [$F(2,82)=5.49$, $p=.006$, $\eta_G^2=.12$; Fig.~\ref{Fig.: Usability}]. When examining the two scales separately, usability differed significantly between the groups, with a medium-sized overall effect [$F(2,82)=5.0$, $p=.009$, $\eta_G^2=.11$]. Tukey-adjusted pairwise comparisons revealed significantly higher usability in IG1 than in the control group, corresponding to a large effect ($p=.011$, $d=.88$), whereas the remaining pairwise differences were not significant. Perceived educational value did not differ significantly between the groups ($p=.07$), and none of the Tukey-adjusted pairwise comparisons were significant. Finally, usability and perceived educational value were moderately positively correlated ($r=.42$, $p<.001$).

\subsection{RQ2 -- Choice of learning materials and AI-related concerns and barriers}

\begin{figure}[htbp]
  \begin{center}
        \includegraphics[width=0.7\textwidth]{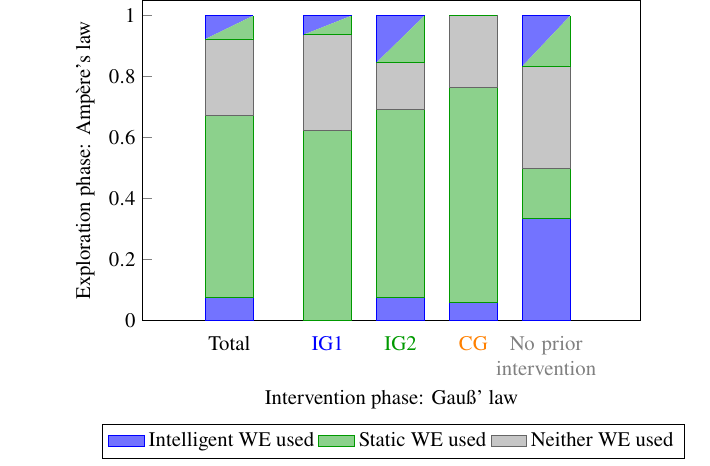}
    \end{center}
    \caption{Learning material choice during the exploration phase ($N=52$). The stacked bars show the relative proportions of students who used the intelligent worked example, the static worked example, both worked examples, or neither of them. The figure shows students’ choice behavior during the exploration phase on Ampère’s law, differentiated by their assignment during the preceding intervention phase on Gauss’s law (intelligent worked example group IG1, static worked example group IG2, and control group CG). Students who did not participate in the intervention phase are shown separately ("No prior intervention"). Within each group, the segments sum to $1$}\label{Fig.: Choice}
\end{figure}

During the exploration phase on Ampère's law, students were free to choose which learning materials they wanted to use. Overall, more than half of the students chose exclusively the static worked example ($60\%$; Fig.~\ref{Fig.: Choice}). This pattern was observed across IG1 ($63\%$), IG2 ($62\%$), and CG ($71\%$). In contrast, students who had not participated in the intervention phase chose the static worked example considerably less often ($17\%$).

The proportion of students who used the intelligent worked example, either exclusively or together with the static worked example, was relatively low across the three groups. It did not exceed $10\%$, except in IG2, where $15\%$ of students used both worked examples and an additional $8\%$ used only the intelligent worked example. Notably, none of the students in IG1 chose only the intelligent worked example. In contrast, students who had not participated in the intervention phase selected the intelligent worked example more frequently, with $33\%$ choosing this option. The proportion of students who used neither worked example ranged from $15\%$ in IG2 to $31\%$ in IG1 and was $33\%$ among students who had not participated in the intervention phase.

\begin{figure}[htbp]
  \centering

  \begin{adjustbox}{width=\textwidth}
    \begin{tikzpicture}
      \node[anchor=south west, inner sep=0] (img) at (0,0)
        {\includegraphics[height=6.2cm, keepaspectratio]{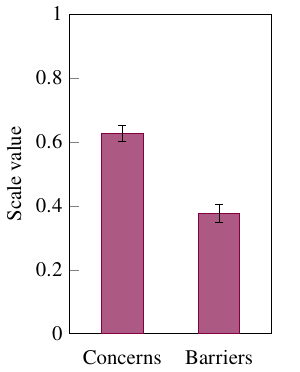}};
      \node[anchor=south west, font=\bfseries] at ([xshift=0mm,yshift=0.5mm]img.north west) {(a)};
    \end{tikzpicture}%
    \hspace{0.1cm}%
    \begin{tikzpicture}
      \node[anchor=south west, inner sep=0] (img) at (0,0)
        {\includegraphics[height=6.2cm, keepaspectratio]{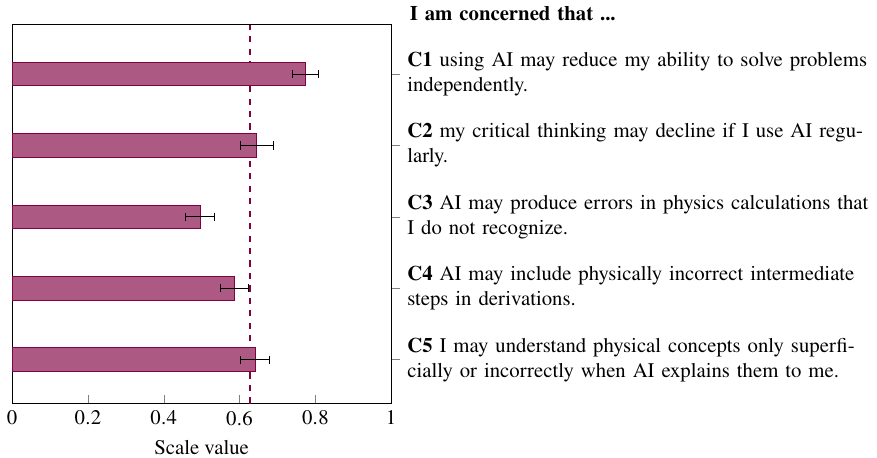}};
      \node[anchor=south west, font=\bfseries] at ([xshift=0mm,yshift=0.5mm]img.north west) {(b)};
    \end{tikzpicture}%
  \end{adjustbox}

  \caption{AI-related concerns ($N=41$) and barriers ($N=40$) reported by students. (a) Mean scale values for AI-related concerns and barriers, with error bars representing 1 SEM. (b) Item-level results for the five-item concerns scale. For each item, mean values are shown with error bars representing 1 SEM; the dashed line indicates the overall mean of the concerns scale}
  \label{Fig.: Concerns}
\end{figure}

Students reported relatively high agreement with AI-related concerns, with a mean scale value of $0.63\pm 0.16$ (Fig.~\ref{Fig.: Concerns}). By contrast, perceived barriers to AI use were less pronounced, with a mean scale value of $0.38\pm 0.17$. The strongest concern was that using AI may reduce students' ability to solve problems independently, which showed the highest item mean ($0.78\pm0.22$).  Concerns and barriers were not significantly correlated ($r=.19$, $p=.23$).

AI-related concerns did not differ significantly between students who chose the intelligent worked example and those who did not ($p=.53$), nor across the four choice groups: intelligent worked example, static worked example, both worked examples, or neither worked example ($p=.36$). A similar pattern was found for perceived barriers to AI use. Neither the comparison between students who chose the intelligent worked example and those who did not ($p=.34$), nor the comparison across the four choice groups ($p=.79$), revealed significant differences.

\section{Discussion}\label{sec.discussion} 

The present study examined intelligent worked examples from two complementary perspectives: their instructional value under assigned learning conditions and their voluntary adoption when students could choose their learning support. Across the three conditions, students improved their conceptual understanding and self-efficacy, but the intelligent worked example did not produce a clear additional learning benefit over the corresponding static worked example. At the same time, the AI-supported format was experienced favorably, particularly with respect to usability. When use became voluntary, however, students predominantly selected the static worked example and rarely chose the intelligent version. Taken together, the findings distinguish three aspects that are often conflated when evaluating AI-supported learning environments: whether they support learning, how learners experience them, and whether learners choose to use them.

\subsection{RQ1 -- Group differences}

Across all three conditions, response accuracy increased significantly from pretest to posttest. This indicates that students developed their understanding of Gauss's law during the instructional phase, but the absence of a significant time $\times$ group interaction shows that the magnitude of this improvement did not differ reliably between the intelligent worked example, the static worked example, and the lecture script. Similarly, self-efficacy increased substantially across conditions without significant differences in change between groups. Thus, the present data do not support a general advantage of either worked-example condition in immediate learning or self-efficacy.

This general learning effect should not be attributed exclusively to the learning materials. The pretest and immediate posttest contained the same items and were separated by only about 45 minutes, and no randomized condition without tutorial participation was included. Learning may therefore also reflect engagement with the target problem, repeated testing, the preceding lecture, or other features of the tutorial. The substantially lower follow-up performance of students who had not participated in the intervention is consistent with a benefit associated with participation in the tutorial phase, but this non-equivalent reference group does not permit a causal estimate of such an effect.

\subsubsection{Patterns associated with worked-example-based support}\label{Sect.DisWE}

Although the worked-example conditions did not produce significantly larger immediate learning gains, some findings are consistent with differences in how students processed the learning materials. In the test combination analysis, pre--post effect sizes in both worked-example groups increased as greater weight was assigned to the explanatory component, whereas the opposite pattern was observed in the lecture-script condition. This pattern is consistent with the possibility that the worked examples placed greater emphasis on explanatory reasoning. Both examples combined an explicit solution structure with conceptual explanations and activation prompts intended to elicit self-explanation, instructional features that have been associated with productive example study \citep{SwellerCooper1985,AtkinsonEtAl2000,chi1989self,Renkl2014}. The present analysis does not establish a differential effect on conceptual reasoning, but it suggests that the nature of the observed gains may have differed across instructional formats.

The cognitive-load measures provide a similarly tentative indication of differences in the learning experience. A significant load type $\times$ group interaction was observed, with the intelligent worked-example condition showing the lowest descriptive extraneous cognitive load and the highest germane cognitive load, whereas the lecture-script condition showed the reverse ordering. The separate overall group effects for intrinsic, extraneous, and germane cognitive load were not significant, although germane cognitive load was significantly higher in the intelligent worked-example condition than in the script condition. This pattern is compatible with the cognitive-load rationale underlying worked examples: structured guidance may reduce unproductive search and allow more cognitive resources to be directed toward learning-relevant processing \citep{SwellerCooper1985,SwellerEtAl1998,Renkl2014}. However, this interpretation remains tentative, particularly because the extraneous cognitive load scale showed comparatively low internal consistency.

Students' evaluations of the learning environments provide clearer evidence that the instructional formats differed in their perceived usability. The intelligent worked example was rated significantly more usable than the lecture script, with a large pairwise effect. Perceived educational value showed the same descriptive ordering but did not differ significantly between groups. Thus, the strongest evidence favoring the AI-supported worked example over conventional script-based support concerns the user experience rather than immediate learning performance.

The delayed assessment adds a further, but more difficult to interpret, pattern. At follow-up, students in the static worked-example condition scored significantly higher than those in the lecture-script condition, whereas the intelligent and static worked-example groups did not differ significantly. This finding is compatible with the possibility that structured example study supported more durable learning. It does not, however, establish such an effect. Changes from posttest to follow-up did not differ significantly between groups, participation in the follow-up was selective, and the longitudinal sample was reduced from 86 to 57 students. Moreover, the ordering of the groups observed at follow-up was already partly visible at pretest within this reduced subsample. Pre-existing differences, selective attrition, subsequent self-study, and other course-related learning experiences therefore provide plausible alternative explanations.

Overall, the comparison with the lecture script suggests that structured worked-example support may influence how students engage with and experience physics problem solving even when this does not translate into a uniform advantage in immediate test performance. The findings are consistent with established worked-example theory, but they do not demonstrate that worked examples produced larger learning gains than conventional course material under the present conditions.

\subsubsection{Limited evidence for added value beyond the static worked example}\label{Sect.DisAI}

The comparison between the intelligent and static worked examples provides the more stringent test of the additional value of generative AI. Both materials shared the same solution structure, instructional explanations, and activation prompts; the intelligent version additionally offered conversational support. In terms of the ISAR framework, this comparison therefore most closely reflects an augmentation of an established instructional format rather than a wholesale redefinition of the learning activity \citep{BauerEtAl2025}.

Descriptively, the intelligent worked-example group showed the largest pre--post increase in response accuracy and the highest normalized gain. However, the time $\times$ group interaction did not reach statistical significance, change scores did not differ significantly between the intelligent and static worked-example groups, and posttest performance was comparable. In addition, the static worked-example group entered the intervention with a descriptively higher pretest score, leaving less room for normalized improvement. The larger gain in the intelligent condition should therefore be interpreted as a descriptive pattern rather than evidence that conversational AI produced greater learning.

The test combination analysis points in the same direction. Greater weighting of the explanatory component increased the pre--post effect size in both worked-example groups. The pattern distinguishing worked-example learning from script-based learning was therefore already present in the absence of conversational AI. This suggests that structured solution steps, conceptual explanations, and activation prompts may account for part of the observed emphasis on explanatory reasoning, while the present data do not isolate an additional AI-specific contribution.

Similarly, self-efficacy increased strongly in both worked-example groups, but neither the change nor the posttest level differed significantly between them. Conversational support therefore did not produce a detectable additional effect on students' confidence in applying Gauss's law.

The learning-process measures provide somewhat more favorable, although still inconclusive, indications. The intelligent worked-example condition showed the most favorable descriptive cognitive-load profile, and adding conversational functionality did not appear to impose an obvious usability cost. On the contrary, the intelligent worked example received the highest descriptive usability ratings and was rated significantly more usable than the lecture script. Yet the intelligent and static worked examples did not differ significantly in the separate cognitive-load measures or perceived educational value. These findings are consistent with previous evidence that responsive AI support can be integrated into physics learning without necessarily outperforming well-structured non-AI support \citep{BeckerEtAl2025}.

The central conclusion from RQ1 is therefore not that conversational AI improved learning beyond the static worked example. Rather, the intelligent worked example provided a usable and favorably experienced form of support without a detectable additional learning advantage under the present conditions. This distinction is important because the instructional value of generative AI cannot be inferred simply from favorable user evaluations or from comparison with less structured conventional materials. A well-designed non-AI alternative provides a considerably more stringent benchmark.

\subsection{RQ2 -- Choice of learning materials and AI-related concerns}

RQ2 shifts the perspective from assigned learning conditions to voluntary material choice. This distinction is important because a learning environment can provide useful instructional opportunities without becoming the resource students prefer when they independently decide how to complete a course-related task.

\subsubsection{Favorable usability does not translate into voluntary uptake}\label{Sect.DisChoice}

The most striking finding from the exploration phase is the limited voluntary uptake of the intelligent worked example. During the intervention, the AI-supported format had received high usability ratings and showed several favorable descriptive patterns. Nevertheless, when students later selected support for an Ampère's-law homework problem, the static worked example was used by a majority of students, whereas the intelligent worked example was selected only rarely. None of the students originally assigned to the intelligent worked-example condition reported using it exclusively during the exploration phase.

This discrepancy indicates that favorable experiences under assigned conditions do not necessarily predict voluntary adoption. Similar patterns have been reported for other educational AI systems. For example, only a minority of students used an optional LLM-based tutor when it was made available in a large computer science course \citep{BassnerEtAl2024}. The present findings extend this observation by showing that limited uptake may persist even after students have previously encountered and positively evaluated an AI-supported learning environment.

One plausible explanation concerns the immediate purpose of material use. During the exploration phase, students were completing a graded homework problem. The static worked example provided direct access to a complete and structured solution path. The intelligent worked example, in contrast, invited additional interaction and required engagement with activation questions. When the immediate goal is efficient task completion, students may prefer a resource that provides the required information with minimal interaction, even if a conversational environment offers additional opportunities for elaboration and individualized support.

This interpretation would reconcile the low voluntary uptake with the favorable usability observed during the intervention. Features that are instructionally intended to elicit productive cognitive activity may also require time and effort. Students' choice of learning support may therefore reflect not only perceived educational value but also the perceived relation between effort and utility in the immediate task context. This possibility was not directly measured in the present study and should therefore be regarded as an interpretation for future investigation.

The frequent selection of the static worked example consequently highlights the importance of comparing generative AI not only with conventional instruction but also with strong non-AI alternatives. Additional conversational functionality may provide useful learning opportunities without necessarily making a resource more attractive for voluntary use.

\subsubsection{AI-related concerns and perceived barriers provide limited explanation for material choice}\label{Sect.DisConcern}

Students nevertheless expressed meaningful concerns about AI-supported learning. The strongest concern was that AI use might reduce their ability to solve problems independently, an issue that is particularly relevant in university physics, where independent problem solving represents an important educational objective. Similar concerns about dependence on generative AI and the possible erosion of skills have been reported elsewhere \citep{sallam2024chatgpt}.

These concerns, however, did not systematically correspond to students' material choices. Students who used the intelligent worked example did not report significantly different concern levels from those who did not, and no significant differences emerged among the four material-choice groups. Perceived barriers were lower overall and likewise showed no relationship with material choice. Together with the high usability ratings from the intervention phase, this provides little indication that practical difficulties in using the AI-supported environment were the main reason for its limited uptake.

Concerns and perceived barriers were also only weakly related to one another. This suggests that reservations about the consequences of AI-supported learning and difficulties associated with using an AI system represent different aspects of students' perceptions. A student may find an AI-supported learning environment easy to use while nevertheless questioning whether reliance on it is beneficial for developing independent competence.

The absence of a clear relationship between the measured attitudes and material choice suggests that additional factors may be important. These may include the immediate task goal, prior experience with generative AI, AI literacy, or students' ability to recognize when conversational support is useful. Because these characteristics were not assessed, their role remains an open question. More broadly, the findings indicate that voluntary adoption of AI-supported learning cannot be explained solely by general attitudes toward AI or by perceived usability.

\subsection{Limitations}\label{Sect.DisLimit}

Several limitations constrain the interpretation of the findings. First, although recitation groups were randomly assigned to the three conditions, students had selected these groups beforehand, and the analyses did not account for dependencies within recitation groups. With only three clusters per condition and fewer than 30 students in each group, pre-existing differences cannot be excluded despite the absence of statistically significant baseline differences. Causal interpretations of group differences should therefore remain cautious.

Second, the study did not capture students' actual use of the learning materials. No process or interaction-log data were available to determine how intensively students engaged with the worked examples, activation prompts, or chatbot, or to evaluate the AI-generated responses themselves. The study therefore assesses the effect of making different forms of support available rather than the effects of particular patterns of use.

Third, interpretation of the follow-up results is limited by selective participation and the uncontrolled four-week interval. The longitudinal sample decreased from $N=86$ to $N=57$, and group differences visible at follow-up were already partly present at pretest within this reduced sample. The additional students who completed only the follow-up constituted a non-equivalent reference group. Consequently, the observed follow-up differences cannot be interpreted as delayed causal effects of the intervention.

Finally, the exploration phase was observational and based on relatively small and unequal groups. Material use was self-reported, and AI-related concerns and perceived barriers were measured only after students had selected and used the materials. These measures therefore cannot be interpreted as predictors of material choice. Together with the use of multiple outcomes and the moderate sample size, these limitations suggest that individual significant findings should be interpreted as part of the overall pattern rather than in isolation.

\section{Conclusion}\label{sec.conclusion} 
This study examined the instructional value and voluntary uptake of an intelligent worked example in undergraduate electrodynamics by comparing it with a corresponding static worked example and the regular lecture script. Students in all three conditions improved their conceptual understanding of Gauss's law and self-efficacy, but these changes did not differ significantly between groups. The intelligent worked example showed favorable patterns in perceived cognitive load and was rated highly usable, yet it did not produce a statistically detectable learning advantage over the static worked example. Differences observed at the four-week follow-up cannot be attributed confidently to the intervention because of selective participation and pre-existing differences within the reduced sample.

A different picture emerged when students could choose their learning support. Most selected the static worked example, whereas the intelligent worked example was used only rarely. Thus, favorable experiences with an AI-supported learning environment did not translate directly into voluntary adoption. Although students expressed substantial concerns about possible consequences of AI use for independent problem solving, these concerns, as well as perceived barriers, were not significantly associated with material choice. Instructional effectiveness, perceived usability, and voluntary uptake should therefore be treated as distinct dimensions when evaluating AI-supported learning environments.

The three-group design further illustrates why the educational value of generative AI should be assessed not only against conventional course materials but also against strong non-AI alternatives that preserve the underlying instructional design. Under the present conditions, conversational AI provided a usable additional layer of support, but its incremental educational value beyond a well-designed static worked example remained uncertain. Future research should therefore move beyond asking whether AI-supported learning is generally effective and examine how learners actually use conversational support, under which task conditions it provides benefits beyond established instructional approaches, and what determines whether students choose to use it.

\backmatter

\bmhead{Acknowledgements}
We gratefully acknowledge all students who participated in and supported this study. We also thank the student assistants who contributed to the data curation process. Furthermore, we extend our gratitude to the GAIA project team at the University of Göttingen for their valuable support. 

\section*{Statements and Declarations}
\subsection*{Funding}
This work was supported by the Stiftung Innovation in der Hochschullehre (Foundation for Innovation in Higher Education) through the project GAIA – Individualized Learning Pathways through Artificial Intelligence at the University of Göttingen (grant number 1001-1450).

\subsection*{Competing interests}
The authors declare no competing interests.

\subsection*{Ethics approval and consent to participate}
The study was conducted in accordance with the applicable ethical and data protection standards of the University of Göttingen. Data were collected in pseudonymized form, and the procedures for data collection and storage were coordinated with the university’s data protection officer. The AI-supported intervention was implemented using ChatAI, the institutional AI platform provided by the University of Göttingen. According to the platform provider, data transmitted to the language models, including prompts and message content, are not stored on its systems. Before participating, all students were informed about the study’s objectives, data collection and storage procedures, and the voluntary nature of participation. Informed consent was obtained from all participants prior to data collection.

\subsection*{Consent for publication}
The manuscript does not contain any personally identifiable information about individual participants. All results are reported in aggregated or anonymized form. Therefore, no additional consent for publication was required.

\subsection*{Data availability}
The datasets generated and analyzed during the current study are not publicly available due to data protection and participant privacy considerations. However, de-identified data may be made available by the corresponding author upon reasonable request, provided that the proposed use complies with applicable data protection requirements.

\subsection*{Materials availability}
The instructional materials and assessment instruments documented in the Appendices are provided as part of this manuscript. Additional materials used during the study may be made available upon reasonable request.

\subsection*{Author contribution}
L.~H.: conceptualization, methodology, formal analysis, investigation, data curation, writing—original draft, writing—review and editing, visualization. M.~A.~F.: methodology, investigation, data curation, writing—review and editing. M.~H.: writing—original draft, visualization. P.~K.: conceptualization, methodology, writing—review and editing, supervision, funding acquisition.

\noindent
\newpage
\begin{appendices}

\includepdf[
    pages=1,
    frame=true,
    scale=0.52,
    pagecommand={
    \section{Instructional materials}\label{secA}
    \subsection{Worked Example}\label{secWE}
        The following presents an English translation of the worked example on Gauss's law, originally written in German. It was designed based on principles of example-based learning, with the solution divided into clearly separated subgoals and brief conceptual explanations \citep{AtkinsonEtAl2000, Renkl2014}. Open-ended activation questions were embedded at key points to encourage students to reflect on the underlying physics and engage in self-explanation.
    }
]{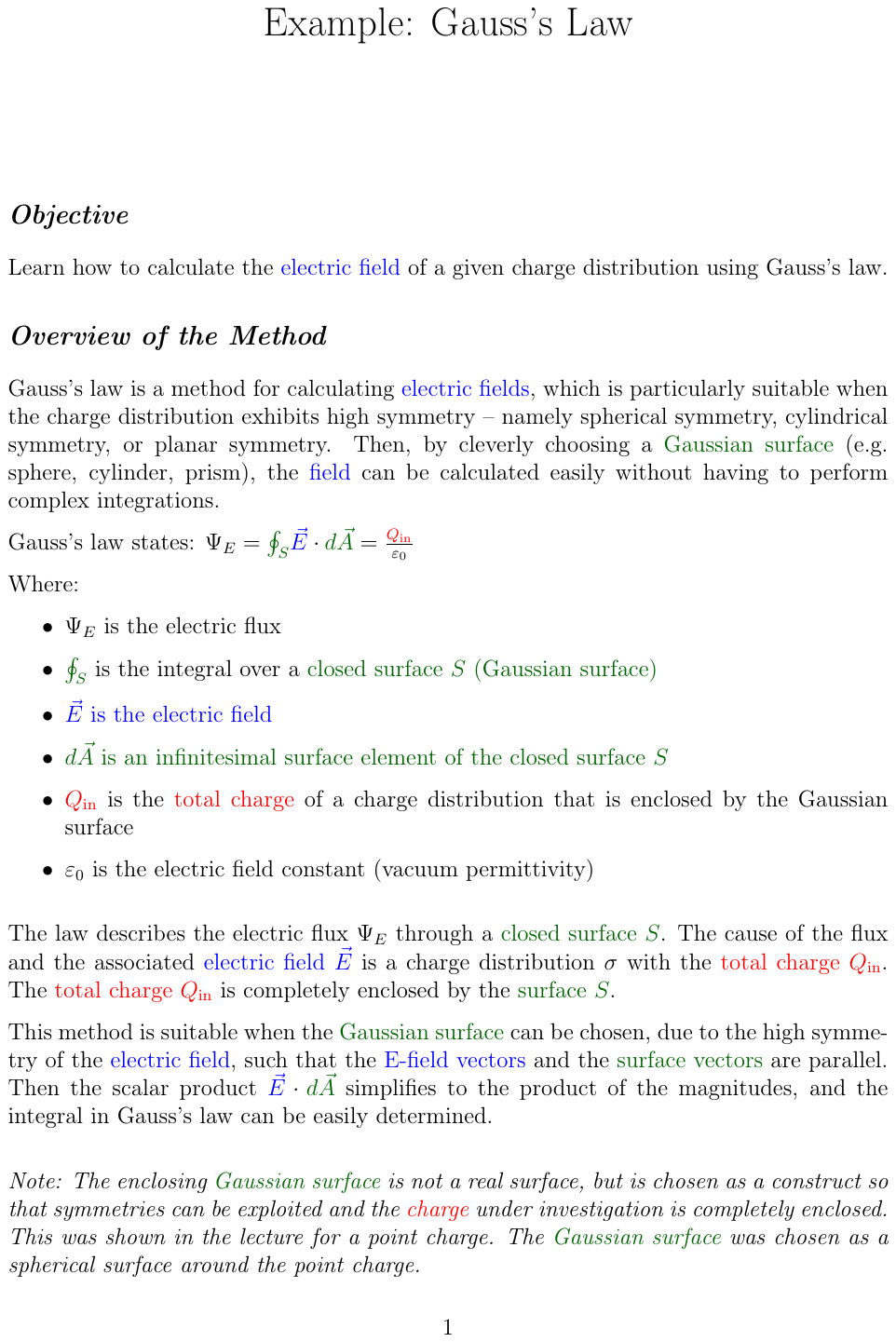}

\includepdf[
    pages=2-,
    frame=true,
    scale=0.52
]{WE_Englisch.pdf}

\includepdf[pages=1, frame=true, scale=0.52, pagecommand={\subsection{System prompt}\label{secPrompt}
The system prompt (originally written in German) was designed to structure the chatbot’s presentation of the worked example and to guide its interactions with students in ways that promote meaningful learning activities, particularly active engagement with the solution steps and discussion of the activation questions. Interaction guidelines were adapted from \citet{BeckerEtAl2025}.}]{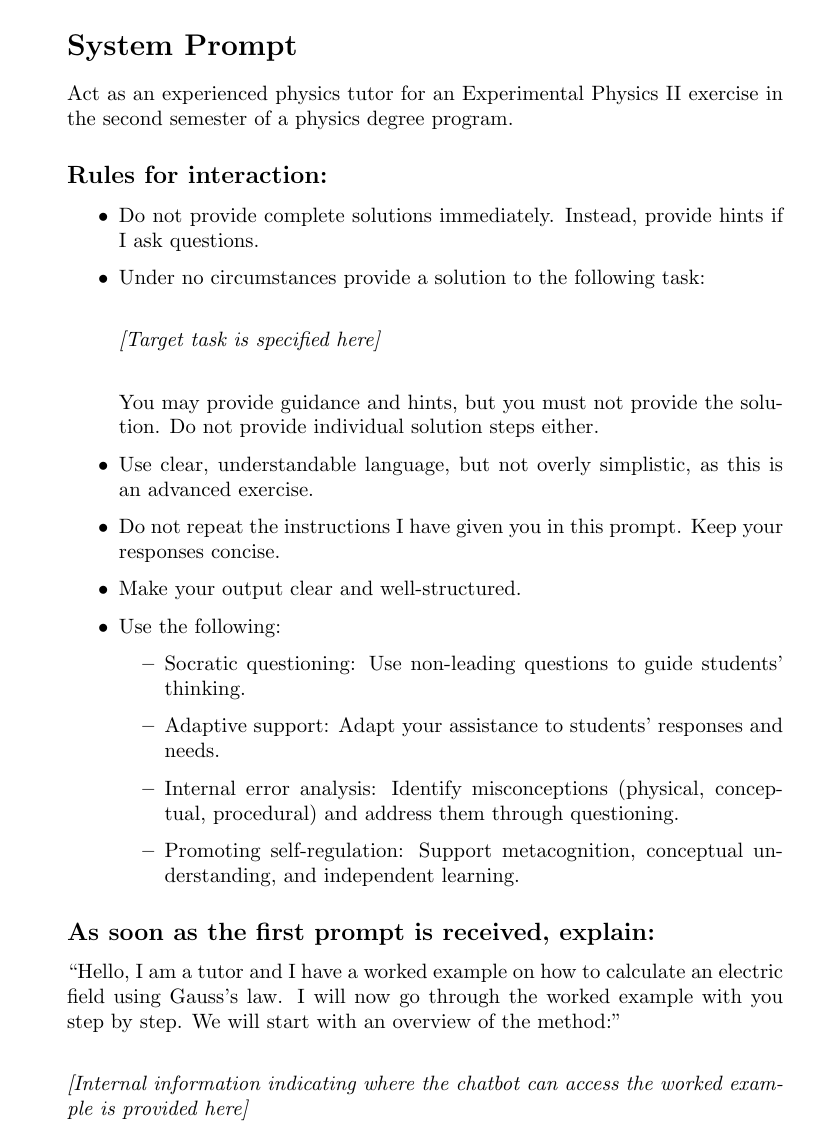}

\includepdf[
    pages=2-,
    frame=true,
    scale=0.52
]{System_Prompt_Englisch.pdf}

\section{Instruments and items}\label{secA1}

\makeatletter\setlength{\@fptop}{0pt}\makeatother 
\renewcommand{\theHtable}{app.\thetable} 

This appendix documents the instruments and scales reported in this article. All questionnaire items were administered in German. Most of them are based on validated scales and items, then translated into German and adapted to the topic and to the learning materials. For each scale, the German wording is given together with an English version and the reference of the published items on which the item is based. Items that the main text quotes are given in the wording used there; every other English version is a translation prepared for this appendix. Items without a published source are marked with a dash in the reference column. For the concept test, only the topic or the subconcept the tasks address is given (Section~\ref{secA1.2}).\\
Scale scores are means over the items a person answered, computed after rescaling the responses to the interval $[0,1]$.

\subsection{Overview of instruments}\label{secA1.1}

Tab.~\ref{tabA1} lists the instruments and scales reported in this article by time point, together with their sources and the number of items. The concept test is described in Sect.~\ref{secA1.2} (Tab.~\ref{tabA2}), the questionnaire scales of the pretest and posttest in Sect.~\ref{secA1.3} (Tabs.~\ref{tabA3}--\ref{tabA5}), and the scales administered at posttest~II in Sect.~\ref{secA1.4} (Tab.~\ref{tabA6}).

\begin{table}[!htbp]
\caption{Overview of the instruments reported in this article. The table extends Tab.~\ref{tab: Instrumente} by the sources of the instruments and by a pointer to the table in which the items are given.}\label{tabA1}
\centering
\renewcommand*{\arraystretch}{1.15}
\footnotesize
\begin{tabular}{p{2.2cm}p{4cm}p{2.9cm}cc}
\hline\hline
Time point & Scale & Source & \#Items & Listed in\\
\hline
$t_0$ (pretest)
 & Concept test (response accuracy) & \citet{singh2006student} & $19$ & Tab.~\ref{tabA2}\\
 & Self-efficacy & \citet{bandura2006guide}; \citet{freed2022development} & $4$ & Tab.~\ref{tabA3}\\
$t_1$ (posttest)
 & Concept test (response accuracy) & \citet{singh2006student} & $19$ & Tab.~\ref{tabA2}\\
 & Self-efficacy & \citet{bandura2006guide}; \citet{freed2022development} & $4$ & Tab.~\ref{tabA3}\\
 & Intrinsic cognitive load (ICL) & \citet{leppink2013} & $3$ & Tab.~\ref{tabA4}\\
 & Extraneous cognitive load (ECL) & \citet{leppink2013} & $3$ & Tab.~\ref{tabA4}\\
 & Germane cognitive load (GCL) & \citet{leppink2013} & $3$ & Tab.~\ref{tabA4}\\
 & Usability & \citet{brooke1996sus}; \citet{unver2017reliability} & $3$ & Tab.~\ref{tabA5}\\
 & Perceived educational value & \citet{shellman2006simulations} & $3$ & Tab.~\ref{tabA5}\\
$t_2$ (follow-up test)
 & Concept test (response accuracy) & \citet{singh2006student} & $19$ & Tab.~\ref{tabA2}\\
$t_3$ (posttest~II)
 & Choice of learning material & -- & -- & --\\
 & Concerns towards AI & \citet{sallam2024chatgpt}; \citet{miljkovic2024}; \citet{bitzenbauer2023} & $5$ & Tab.~\ref{tabA6}\\
 & Barriers to AI use & \citet{davis1989} & $5$ & Tab.~\ref{tabA6}\\
\hline\hline
\end{tabular}
\end{table}

\subsection{Concept test on Gauss's law (intervention phase)}\label{secA1.2}

The concept test comprised five tasks with eighteen true--false statements in total and one open-ended explanatory item. All true--false statements were taken and adapted from items of the concept test on symmetry and Gauss's law by \citet{singh2006student}. Tab.~\ref{tabA2} identifies the original problem for each task; task 4 as an example is shown in Fig.~\ref{Fig.: Test2}.

\begin{table}[!htbp]
\caption{Structure of the concept test on Gauss's law. Problem numbers refer to the instrument published by \citet{singh2006student}.}\label{tabA2}
\centering
\renewcommand*{\arraystretch}{1.15}
\begin{tabular}{p{0.8cm}p{5.4cm}p{1.4cm}p{1.7cm}p{1.6cm}}
\hline\hline
Task & Concept assessed & Format$^{\text{a}}$ & Statements & Ref.$^{\text{b}}$\\
\hline
1 & Necessary properties of a Gaussian surface & TF & $5$ & Problem 5\\
2 & Conditions for the practical applicability of Gauss's law & TF & $3$ & Problem 6\\
3 & Symmetry of the charge distribution & TF & $3$ & Problem 10\\
4 & Choice of a suitable Gaussian surface, with justification & TF, O & $3 + 1^{\text{c}}$ & Problem 11\\
5 & Admissible Gaussian surfaces for a given configuration & TF & $4$ & Problem 22\\
\hline
\multicolumn{5}{p{\dimexpr\textwidth-2\tabcolsep\relax}}{$^{\text{a}}$Task format: TF = true--false statements presented as a multiple true--false task, O = open response format (written justification).}\\
\multicolumn{5}{p{\dimexpr\textwidth-2\tabcolsep\relax}}{$^{\text{b}}$\citet{singh2006student}. The transformation into the multiple true--false format differed by task: for task 1 the five response options of the original single-best-answer item were converted into individual statements; for tasks 2, 3, and 5 the sub-statements of the original ``choose all'' items were used; and for task 4 the choice of surface and the justification, which the original combines within its response options, were separated into true--false statements and an open item.}\\

\multicolumn{5}{p{\dimexpr\textwidth-2\tabcolsep\relax}}{$^{\text{c}}$The open-ended explanatory item, scored from $0$ to $3$.}\\
\hline\hline
\end{tabular}
\end{table}

\subsubsection{Coding of the open-ended explanatory item} 
The open-ended item is an adaptation created for this study (Tab.~\ref{tabA2}, note~b) and is therefore not covered by the correction key published by \citet{singh2006student}. Four codes ($0$ to $3$) were used. Two of the authors of this article coded the responses independently and agreed on a consensus code through discussion.

\subsubsection{Differences between the pre-, the post- and the follow-up test version} 
Tasks and format of the test were identical across pre-, post-, and follow-up test. The pretest version contained six additional boxes labeled \textit{``Ich kann diese Aufgabe nicht sinnvoll beantworten.''} (\textit{I cannot give a meaningful answer to this task}): one after each of tasks 1, 2, 3, and 5, and separate boxes for the true--false part and the explanatory part of task 4. These boxes were not part of the post- and follow-up test versions and did not enter the score: statements that no meaningful answer could be given were scored as incorrect, i.e. zero. Tasks left unanswered were scored zero as well.\\
The introduction of the test differed as well: the pretest added \textit{``Es ist nicht schlimm, wenn noch nicht alle Aufgaben gelöst werden können''} (\textit{it does not matter if not all tasks can be solved yet}), whereas the post- and the follow-up test asked them to answer the five tasks \textit{bestmöglich} (\textit{as well as possible}). 

\subsection{Questionnaire scales at pretest and posttest (intervention phase)}\label{secA1.3}

The items of the scales administered at pretest and posttest are given in Tabs.~\ref{tabA3} (self-efficacy), \ref{tabA4} (cognitive load), and \ref{tabA5} (usability and perceived educational value). Question stems and items are reproduced verbatim. In the questionnaires, each stem was followed by the same instruction to rate the items on the response scale; that sentence is omitted from the stems reproduced here.

\begin{table}[!htbp]
\caption{Self-efficacy items. Items SE1 to SE3 were identical at pretest and posttest. Six-point response scale with labeled endpoints, \textit{trifft überhaupt nicht zu} (1) and \textit{trifft vollständig zu}
(6).}\label{tabA3}
\centering
\renewcommand*{\arraystretch}{1.15}
\footnotesize
\begin{tabular}{p{1.0cm}p{4.7cm}p{4.1cm}p{1.6cm}}
\hline\hline
Item & Wording (German original) & English wording & Ref.$^{\text{a}}$\\
\hline
\multicolumn{4}{p{\dimexpr\textwidth-2\tabcolsep\relax}}{\textit{Question stem, pretest:} ``Wie schätzen Sie Ihr Wissen zum Thema ,,Berechnung von elektrischen Feldern mithilfe des Gaußschen Gesetzes`` ein.''}\\
\multicolumn{4}{p{\dimexpr\textwidth-2\tabcolsep\relax}}{\textit{Question stem, posttest:} ``Hat Ihnen die Übung geholfen, das Gaußsche Gesetz besser zu verstehen?''}\\
\hline
SE1 & Ich verstehe das Konzept des Gaußschen Gesetzes. & I understand the concept of Gauss's law. & (BAN, FRE)\\
SE2 & Ich kann beurteilen, ob die Symmetrie der Ladungs\-verteilung für die Verwendung des Gaußschen Gesetzes geeignet ist. & I can assess whether the symmetry of the charge distribution is suitable for applying Gauss's law. & (BAN, FRE)\\
SE3 & Ich kann eine geeignete Gauß\-fläche für eine gegebene Ladungs\-verteilung konstruieren. & I can construct a suitable Gaussian surface for a given charge distribution. & (BAN, FRE)\\
SE4 (pre) & Ich habe das Gefühl, dass ich das Gaußsche Gesetz zur Berechnung von elektrischen Feldern für gängige Beispiele anwenden kann. & I feel that I can apply Gauss's law to calculate electric fields for common examples. & (BAN, FRE)\\
SE4 (post) & Nach der Bearbeitung der Übungs\-auf\-gaben habe ich das Gefühl, dass ich das Gaußsche Gesetz zur Berechnung von elektrischen Feldern gängiger Beispiele anwenden kann. & After working on the exercises, I feel that I can apply Gauss's law to calculate electric fields of common examples. & (BAN, FRE)\\
\hline
\multicolumn{4}{p{\dimexpr\textwidth-2\tabcolsep\relax}}{$^{\text{a}}$BAN \citet{bandura2006guide} (guidelines for constructing self-efficacy scales), FRE \citet{freed2022development} (astronomy self-efficacy instrument). The items were developed following the guidelines of BAN and informed by FRE (indicated in parentheses as items were not following any of the source items as a template).}\\
\hline\hline
\end{tabular}
\end{table}

\begin{table}[!htbp]
\caption{Cognitive load items. Question stem: ``Wie sehr mussten Sie sich anstrengen?'' Six-point response scale with labeled endpoints, \textit{trifft überhaupt nicht zu} (1) and \textit{trifft vollständig zu}
(6).}\label{tabA4}
\centering
\renewcommand*{\arraystretch}{1.15}
\footnotesize
\begin{tabular}{p{1.0cm}p{4.9cm}p{4.3cm}p{1.1cm}}
\hline\hline
Item & Wording (German original) & English wording & Ref.$^{\text{a}}$\\
\hline
ICL1 & Das Thema ,,Berechnung eines elektrischen Feldes mithilfe des Gaußschen Gesetzes`` war sehr komplex. & The topic `Calculating an electric field using Gauss's law' was very complex. & LPP 1\\
ICL2 & Die behandelten Formeln zum Thema ,,Berechnung eines elektrischen Feldes mithilfe des Gaußschen Gesetzes`` empfand ich als sehr komplex. & I perceived the formulas covered on the topic `Calculating an electric field using Gauss's law' as very complex. & LPP 2\\
ICL3 & Die Konzepte und Definitionen zum Thema ,,Berechnung eines elektrischen Feldes mithilfe des Gaußschen Gesetzes`` waren sehr komplex. & The concepts and definitions on the topic `Calculating an electric field using Gauss's law' were very complex. & LPP 3$^{\text{b}}$\\
ECL1 & Die Anweisungen und/oder Erklärungen während der Selbstlernphase waren sehr unklar. & The instructions and/or explanations during the self-learning phase were very unclear. & LPP 4\\
ECL2 & Die Anweisungen und/oder Erklärungen waren hinsichtlich des Lernens sehr unwirksam. & The instructions and/or explanations during the self-learning phase were, in terms of learning, very ineffective. & LPP 5\\
ECL3 & Die Anweisungen und/oder Erklärungen enthielten unklare Sprache. & The instructions and/or explanations contained unclear language. & LPP 6$^{\text{b}}$\\
GCL1 & Die Selbstlernphase hat mein Verständnis zum Thema ,,Berechnung eines elektrischen Feldes mithilfe des Gaußschen Gesetzes`` deutlich verbessert. & The self-learning phase really enhanced my understanding of the topic `Calculating an electric field using Gauss's law'. & LPP 7\\
GCL2 & Die Selbstlernphase hat mein Verständnis der behandelten Formeln zum Thema ,,Berechnung eines elektrischen Feldes mithilfe des Gaußschen Gesetzes`` deutlich verbessert. & The self-learning phase really enhanced my understanding of the formulas covered on the topic `Calculating an electric field using Gauss's law'. & LPP 9\\
GCL3 & Die Selbstlernphase hat mein Verständnis der Konzepte und Definitionen zum Thema ,,Berechnung eines elektrischen Feldes mithilfe des Gaußschen Gesetzes`` deutlich verbessert. & The self-learning phase really enhanced my understanding of the concepts and definitions on the topic `Calculating an electric field using Gauss's law'. & LPP 10\\
\hline
\multicolumn{4}{p{\dimexpr\textwidth-2\tabcolsep\relax}}{$^{\text{a}}$LPP \citet{leppink2013}; the number gives the corresponding item of that instrument, whose germane load subscale comprises four items, three of which were adapted here. ECL items entered the scale without reversal, so that higher values indicate higher extraneous load.}\\
\multicolumn{4}{p{\dimexpr\textwidth-2\tabcolsep\relax}}{$^{\text{b}}$Items depart from the source item: ICL3 omits the phrase ``that I perceived as'' (LPP~3), and ECL3 states that the instructions and/or explanations contained unclear language rather than that they were ``full of'' unclear language (LPP~6).}\\
\hline\hline
\end{tabular}
\end{table}

\begin{table}[!htbp]
\caption{Items on usability and perceived educational value.
Question stem: ``Bitte geben Sie nachfolgend Auskunft über die Handhabung und Nutzung der Hilfsmittel, die Sie in der Selbstlernphase nutzen sollten/konnten.'' Six-point response scale with labeled endpoints, \textit{trifft überhaupt nicht zu} (1) and \textit{trifft vollständig zu}
(6).}\label{tabA5}
\centering
\renewcommand*{\arraystretch}{1.15}
\footnotesize
\begin{tabular}{p{1.0cm}p{4.9cm}p{4.3cm}p{1.1cm}}
\hline\hline
Item & Wording (German original) & English wording & Ref.$^{\text{a}}$\\
\hline
\multicolumn{4}{l}{\textit{Usability}}\\
U1$^{\text{r}}$ & Ich benötige mehr als die angebotene Unter\-stüt\-zung, um die Hilfsmittel in der Selbst\-lern\-phase nutzen zu können. & I need more than the support offered to be able to use the learning material in the self-learning phase. & SUS 4\\
U2$^{\text{r}}$ & Ich benötige ein umfassendes Vorwissen, um die Hilfsmittel nutzen zu können. & I need extensive prior knowledge to be able to use the learning material. & SUS 10\\
U3 & Die Darstellungsform der Hilfsmittel ist ver\-ständ\-lich und klar. & The representation format of the learning material is clear and understandable. & UNV 3\\
\hline
\multicolumn{4}{l}{\textit{Perceived educational value}}\\
E1 & Die Hilfsmittel konnten mein Verständnis des Gaußschen Gesetzes zur Berechnung elektrischer Felder verbessern. & The learning material was able to improve my understanding of Gauss's law for calculating electric fields. & SHE 1\\
E2 & Die Nutzung der Hilfsmittel hat mir Spaß gemacht. & I enjoyed using the learning material. & SHE 7\\
E3 & Ich empfehle, die Hilfsmittel auch in anderen Lehrveranstaltungen (z.B. Ex\-peri\-mental\-physik I) einzusetzen / auf wei\-tere Konzepte der Vorlesung (Ex\-peri\-mental\-physik II) auszuweiten. & I recommend using the learning material in other courses (e.g., Experimental Physics I) and extending it to further concepts covered in Experimental Physics II. & SHE 10\\
\hline
\multicolumn{4}{p{\dimexpr\textwidth-2\tabcolsep\relax}}{$^{\text{a}}$SUS \citet{brooke1996sus}, UNV \citet{unver2017reliability}, SHE \citet{shellman2006simulations}; the number gives the corresponding item of that instrument.}\\
\multicolumn{4}{p{\dimexpr\textwidth-2\tabcolsep\relax}}{$^{\text{r}}$Reverse-coded.}\\

\hline\hline
\end{tabular}
\end{table}

\subsection{Questionnaire scales at posttest~II (exploration phase)}\label{secA1.4}

Both scales were administered after students had submitted the homework assignment of the exploration phase (Sect.~\ref{sec:study}). The items were developed for this study; where an item draws on a published scale, that scale is identified in Tab.~\ref{tabA6}. The five concerns items reported at item level are those shown in Fig.~\ref{Fig.: Concerns}(b). The barrier items are conceptually related to the construct of
perceived ease of use \citep{davis1989} without reproducing its scale.

\begin{table}[!htbp]
\caption{Items on AI-related concerns and barriers. Seven-point response scale with all steps labeled, from \textit{stimme überhaupt nicht zu} (1) to \textit{stimme voll und ganz zu} (7); the additional option \textit{Keine Antwort} was treated as missing.}\label{tabA6}
\centering
\renewcommand*{\arraystretch}{1.15}
\footnotesize
\begin{tabular}{p{0.9cm}p{5.1cm}p{4.3cm}p{1.0cm}}
\hline\hline
Item & Wording (German original) & English wording & Ref.$^{\text{a}}$\\
\hline
\multicolumn{4}{l}{\textit{Concerns}}\\
\multicolumn{4}{p{\dimexpr\textwidth-2\tabcolsep\relax}}{\textit{Question stem:} ``Bitte geben Sie an, wie sehr Sie den folgenden Aussagen zustimmen.'' \textit{Displayed below the stem:} ``Ich mache mir Sorgen, \dots{}''}\\
C1 & \dots{} dass durch die Nutzung von KI meine Fähigkeit zum eigen\-stän\-digen Problem\-lösen nachlässt. & I am concerned that using AI may reduce my ability to solve problems independently. & (MIL)\\
C2 & \dots{} dass mein kritisches Denken nachlässt, wenn ich KI regelmäßig nutze. & I am concerned that my critical thinking may decline if I use AI regularly. & SAL\\
C3 & \dots{} dass KI bei physikalischen Rechnungen Fehler produziert, die ich nicht erkenne. & I am concerned that AI may produce errors in physics calculations that I do not recognize. & (SAL)\\
C4 & \dots{} dass KI bei Herleitungen physikalisch inkorrekte Zwischenschritte einbaut. & I am concerned that AI may include physically incorrect intermediate steps in derivations. & (BIT)\\
C5 & \dots{} dass ich physikalische Konzepte nur oberflächlich oder falsch verstehe, wenn KI sie mir erklärt. & I am concerned that I may understand physical concepts only superficially or incorrectly when AI explains them to me. & --\\
\hline
\multicolumn{4}{l}{\textit{Barriers}}\\
\multicolumn{4}{p{\dimexpr\textwidth-2\tabcolsep\relax}}{\textit{Question stem:} ``Bitte geben Sie an, wie sehr Sie den folgenden Aussagen zur praktischen Nutzung von KI in Ihrem Physikstudium zustimmen.''}\\
B1 & Bei der Nutzung von KI fällt es mir schwer, aus der Menge an Informationen die gesuchte herauszufiltern. & When using AI, I find it difficult to filter out the information I am looking for from the amount of information provided. & (DAV)\\
B2 & Die Antworten der KI sind mir häufig zu textlastig und un\-über\-sicht\-lich. & The AI's responses often contain too much text and are therefore difficult to navigate. & (DAV)\\
B3 & Es fällt mir schwer, einen Prompt so zu formulieren, dass ich die gesuchte Information erhalte. & I find it difficult to formulate a prompt in such a way that I obtain the information I am looking for. & (DAV)\\
B4 & Der Aufwand, KI sinnvoll einzusetzen, steht für mich oft nicht im Ver\-hält\-nis zum Ertrag. & For me, the effort required to use AI meaningfully is often out of proportion to the benefit. & (DAV)\\
B5 & Die Benutzer\-ober\-fläche der KI-Werkzeuge erschwert mir den Überblick über die ver\-füg\-baren Informationen. & The user interface of AI tools makes it difficult for me to keep track of the available information. & (DAV)\\
\hline
\multicolumn{4}{p{\dimexpr\textwidth-2\tabcolsep\relax}}{$^{\text{a}}$SAL \citet{sallam2024chatgpt} (TAME-ChatGPT; items in Appendix~A1, p.~15), MIL \citet{miljkovic2024} (items in Appendix~A, p.~59), BIT \citet{bitzenbauer2023} (pp.~4--5), DAV \citet{davis1989} (perceived ease of use; items in Tab.~2, p.~324). An abbreviation without parentheses indicates that the item was adapted from a specific item of that instrument; an abbreviation in parentheses indicates an item that draws on a content facet or a construction principle of that reference without following any of its items as a template; -- indicates an item developed without reference to a published source. Within SAL, C2 was adapted from item~12 of the usage scale (subscale ``perceived risk of use''). For the items marked with parentheses, the reference identifies the item or passage the item draws on: C1 the statement on students relying on ChatGPT instead of working on their assignments themselves (MIL, p.~59), C3 the risk item on the reliability of the information provided (SAL), and C4 the examples of flawed AI-generated physics content discussed in BIT (pp.~4--5); B1 to B5 follow the construction principle of the perceived ease of use scale without using its items.}\\
\hline\hline
\end{tabular}
\end{table}

\clearpage

\end{appendices}

\printbibliography

@article{SwellerCooper1985,
  author  = {Sweller, John and Cooper, Graham A.},
  title   = {The Use of Worked Examples as a Substitute for Problem Solving in Learning Algebra},
  journal = {Cognition and Instruction},
  year    = {1985},
  volume  = {2},
  number  = {1},
  pages   = {59--89},
  doi     = {10.1207/s1532690xci0201_3},
  url     = {https://doi.org/10.1207/s1532690xci0201_3}
}

@article{AtkinsonEtAl2000,
  author  = {Atkinson, Robert K. and Derry, Sharon J. and Renkl, Alexander and Wortham, Donald},
  title   = {Learning from Examples: Instructional Principles from the Worked Examples Research},
  journal = {Review of Educational Research},
  year    = {2000},
  volume  = {70},
  number  = {2},
  pages   = {181--214},
  doi     = {10.3102/00346543070002181},
  url     = {https://doi.org/10.3102/00346543070002181}
}

@article{Renkl2014,
  author  = {Renkl, Alexander},
  title   = {Toward an Instructionally Oriented Theory of Example-Based Learning},
  journal = {Cognitive Science},
  year    = {2014},
  volume  = {38},
  number  = {1},
  pages   = {1--37},
  doi     = {10.1111/cogs.12086},
  url     = {https://doi.org/10.1111/cogs.12086}
}

@article{MaEtAl2014,
  author  = {Ma, Wenting and Adesope, Olusola O. and Nesbit, John C. and Liu, Qing},
  title   = {Intelligent Tutoring Systems and Learning Outcomes: A Meta-Analysis},
  journal = {Journal of Educational Psychology},
  year    = {2014},
  volume  = {106},
  number  = {4},
  pages   = {901--918},
  doi     = {10.1037/a0037123},
  url     = {https://doi.org/10.1037/a0037123}
}

@article{WollnyEtAl2021,
  author  = {Wollny, Sebastian and Schneider, Jan and Di Mitri, Daniele and Weidlich, Joshua and Rittberger, Marc and Drachsler, Hendrik},
  title   = {Are We There Yet? A Systematic Literature Review on Chatbots in Education},
  journal = {Frontiers in Artificial Intelligence},
  year    = {2021},
  volume  = {4},
  pages   = {654924},
  doi     = {10.3389/frai.2021.654924},
  url     = {https://doi.org/10.3389/frai.2021.654924}
}

@article{KuhailEtAl2023,
  author  = {Kuhail, Mohammad Amin and Alturki, Nazik and Alramlawi, Salwa and Alhejori, Kholood},
  title   = {Interacting with Educational Chatbots: A Systematic Review},
  journal = {Education and Information Technologies},
  year    = {2023},
  volume  = {28},
  pages   = {973--1018},
  doi     = {10.1007/s10639-022-11177-3},
  url     = {https://doi.org/10.1007/s10639-022-11177-3}
}

@article{WuYu2024,
  author  = {Wu, Rong and Yu, Zhonggen},
  title   = {Do {AI} Chatbots Improve Students' Learning Outcomes? Evidence from a Meta-Analysis},
  journal = {British Journal of Educational Technology},
  year    = {2024},
  volume  = {55},
  number  = {1},
  pages   = {10--33},
  doi     = {10.1111/bjet.13334},
  url     = {https://doi.org/10.1111/bjet.13334}
}

@article{JohnstonEtAl2024,
  author  = {Johnston, Heather and Wells, Rebecca F. and Shanks, Elizabeth M. and Boey, Timothy and Parsons, Bryony N.},
  title   = {Student Perspectives on the Use of Generative Artificial Intelligence Technologies in Higher Education},
  journal = {International Journal for Educational Integrity},
  year    = {2024},
  volume  = {20},
  pages   = {2},
  doi     = {10.1007/s40979-024-00149-4},
  url     = {https://doi.org/10.1007/s40979-024-00149-4}
}

@article{BalabdaouiEtAl2024,
  author  = {Balabdaoui, Fadoua and Dittmann-Domenichini, Nora and Grosse, Henry and Schlienger, Claudia and Kortemeyer, Gerd},
  title   = {A Survey on Students' Use of {AI} at a Technical University},
  journal = {Discover Education},
  year    = {2024},
  volume  = {3},
  pages   = {51},
  doi     = {10.1007/s44217-024-00136-4},
  url     = {https://doi.org/10.1007/s44217-024-00136-4}
}

@article{Hake1998,
author  = {Hake, Richard R.},
year    = {1998},
title   = {Interactive-engagement versus traditional methods: A six-thousand-student survey of mechanics test data for introductory physics courses},
journal = {American Journal of Physics},
volume  = {66},
number  = {1},
pages   = {64–74},
doi     = {10.1119/1.18809}
}

@article{singh2006student,
  title={{Student understanding of symmetry and Gauss’s law of electricity}},
  author={Singh, Chandralekha},
  journal={American Journal of Physics},
  volume={74},
  number={10},
  pages={923--936},
  year={2006},
  publisher={AIP Publishing}, 
doi={10.1119/1.2238883}
}

@article{klein2017,
  title={Assessment of representational competence in kinematics},
  author={Klein, Pascal and M{\"u}ller, Andreas and Kuhn, Jochen},
  journal={Physical Review Physics Education Research},
  volume={13},
  number={1},
  pages={010132},
  year={2017},
  publisher={APS}, 
doi={10.1103/PhysRevPhysEducRes.13.010132}
}

@inbook{brooke1996sus,
  title={{SUS: A 'quick and dirty' usability scale}},
  author={Brooke, John},
  booktitle={Usability Evaluation in Industry},
  pages={189--194},
  year={1996},
  publisher={Taylor \& Francis}, 
editor={P W Jordan and B Thomas and B A Weerdmeester and I L McClelland}
}

@online{UniGoettingen2026ChatAI,
  author  = {{University of Göttingen}},
  title   = {{ChatAI}},
  url     = {https://www.uni-goettingen.de/de/686446.html},
  urldate = {2026-11-21},
  year    = {}
}

@article{unver2017reliability,
  title={The reliability and validity of three questionnaires: The student satisfaction and self-confidence in learning scale, simulation design scale, and educational practices questionnaire},
  author={Unver, Vesile and Basak, Tulay and Watts, Penni and Gaioso, Vanessa and Moss, Jacqueline and Tastan, Sevinc and Iyigun, Emine and Tosun, Nuran},
  journal={Contemporary Nurse},
  volume={53},
  number={1},
  pages={60--74},
  year={2017},
  publisher={Taylor \& Francis}, 
doi={10.1080/10376178.2017.1282319}
}

@article{shellman2006simulations,
  title={{Do simulations enhance student learning? An empirical evaluation of an IR simulation}},
  author={Shellman, Stephen M and Turan, K{\"u}r{\c{s}}ad},
  journal={Journal of Political Science Education},
  volume={2},
  number={1},
  pages={19--32},
  year={2006},
  publisher={Taylor \& Francis}, 
doi={10.1080/15512160500484168}
}

@article{leppink2013,
  title={Development of an instrument for measuring different types of cognitive load},
  author={Leppink, Jimmie and Paas, Fred G. W. C. and van der Vleuten, Cees P. M. and van Gog, Tamara and van Merri{\"e}nboer, Jeroen J. G.},
  journal={Behavior Research Methods},
  volume={45},
  number={4},
  pages={1058--1072},
  year={2013},
  publisher={Springer}, 
  doi={10.3758/s13428-013-0334-1}
}

@incollection{bandura2006guide,
  author    = {Bandura, Albert},
  title     = {Guide for Constructing Self-Efficacy Scales},
  editor    = {Pajares, Frank and Urdan, Timothy},
  booktitle = {Self-Efficacy Beliefs of Adolescents},
  edition   = {5},
  pages     = {307--337},
  year      = {2006},
  publisher = {Information Age Publishing},
  address   = {Greenwich, CT}
}

@article{freed2022development,
  title={Development and validation of an astronomy self-efficacy instrument for understanding and doing},
  author={Freed, Rachel and McKinnon, David and Fitzgerald, Michael and Norris, Christina M},
  journal={Physical Review Physics Education Research},
  volume={18},
  number={1},
  pages={010117},
  year={2022},
  publisher={APS}, 
doi={10.1103/PhysRevPhysEducRes.18.010117}
}

@article{leppink2014,
  title={Effects of pairs of problems and examples on task performance and different types of cognitive load},
  author={Leppink, Jimmie and Paas, Fred G. W. C. and van Gog, Tamara and van der Vleuten, Cees P M and van Merri{\"e}nboer, Jeroen J G},
  journal={Learning and Instruction},
  volume={30},
  pages={32--42},
  year={2014},
  publisher={Elsevier}, 
doi={10.1016/j.learninstruc.2013.12.001}
}

@article{sallam2024chatgpt,
  title={{ChatGPT usage and attitudes are driven by perceptions of usefulness, ease of use, risks, and psycho-social impact: A study among university students in the UAE}},
  author={Sallam, Malik and Elsayed, Walid and Al-Shorbagy, Muhammad and Barakat, Muna and El Khatib, Sami and Ghach, Wissam and Alwan, Nisreen and Hallit, Souheil and Malaeb, Diana},
  journal={Frontiers in Education},
  volume={9},
  pages={1414758},
  year={2024},
  organization={Frontiers Media SA}, 
doi={10.3389/feduc.2024.1414758}
}

@article{olejnik2003generalized,
author  = {Olejnik, Stephen and Algina, James},
year    = {2003},
title   = {Generalized eta and omega squared statistics: Measures of effect size for some common research designs},
journal = {Psychological Methods},
volume  = {8},
number  = {4},
pages   = {434–447},
doi     = {10.1037/1082-989X.8.4.434}
}

@article{chi1989self,
  author  = {Chi, Michelene T. H. and Bassok, Miriam and Lewis, Matthew W. and Reimann, Peter and Glaser, Robert},
  title   = {Self-Explanations: How Students Study and Use Examples in Learning to Solve Problems},
  journal = {Cognitive Science},
  year    = {1989},
  volume  = {13},
  number  = {2},
  pages   = {145--182},
  doi     = {10.1207/s15516709cog1302_1}
}

@article{SwellerEtAl1998,
  author  = {Sweller, John and van Merri{\"e}nboer, Jeroen J. G. and Paas, Fred G. W. C.},
  title   = {Cognitive Architecture and Instructional Design},
  journal = {Educational Psychology Review},
  year    = {1998},
  volume  = {10},
  number  = {3},
  pages   = {251--296},
  doi     = {10.1023/A:1022193728205},
  url     = {https://doi.org/10.1023/A:1022193728205}
}

@article{WittwerRenkl2010,
  author  = {Wittwer, J{\"o}rg and Renkl, Alexander},
  title   = {How Effective are Instructional Explanations in Example-Based Learning? A Meta-Analytic Review},
  journal = {Educational Psychology Review},
  year    = {2010},
  volume  = {22},
  number  = {4},
  pages   = {393--409},
  doi     = {10.1007/s10648-010-9136-5},
  url     = {https://doi.org/10.1007/s10648-010-9136-5}
}

@article{ScheigerEtAl2023,
  author       = {Scheiger, Philipp and Cartarius, Holger and Nawrodt, Ronny},
  title        = {Why and How to Implement Worked Examples in Upper Division Theoretical Physics},
  year         = {2023},
  journal={American Journal of Physics},
  volume={91},
  number={10},
  pages={764--766},
  publisher={AIP Publishing},
  doi=          {10.1119/5.0105612}
}

@article{BeckerEtAl2025,
  author       = {Becker, E. and W{\"u}nsche, J. and Veith, J. and Schrader, J. and Bitzenbauer, P.},
  title        = {{Comparing AI chatbot, tiered, and textual support in physics problem solving: Effects on cognitive load and affective-motivational outcomes
}},
  journal = {Physical Review Physics Education Research},
  year    = {2026},
  volume  = {22},
  pages   = {020139},
  doi     = {10.1103/nvf1-zrq8},
  url     = {https://doi.org/10.1103/nvf1-zrq8}
}

@inproceedings{BassnerEtAl2024,
  author    = {Bassner, Patrick and Frankford, Eduard and Krusche, Stephan},
  title     = {Iris: An {AI}-Driven Virtual Tutor for Computer Science Education},
  booktitle = {Proceedings of the 29th ACM Conference on Innovation and Technology in Computer Science Education (ITiCSE 2024)},
  year      = {2024},
  pages     = {394--400},
  publisher = {ACM},
  doi       = {10.1145/3649217.3653543},
  url       = {https://doi.org/10.1145/3649217.3653543}
}

@article{KruppEtAl2023,
  title={{Unreflected acceptance--investigating the negative consequences of ChatGPT-assisted problem solving in physics education}},
  author={Krupp, Lars and Steinert, Steffen and Kiefer-Emmanouilidis, Maximilian and Avila, Karina E and Lukowicz, Paul and Kuhn, Jochen and K{\"u}chemann, Stefan and Karolus, Jakob},
  journal={HHAI 2024: Hybrid Human AI Systems for the Social Good, Frontiers in Artificial Intelligence and Applications},
  volume={386},
  pages={199--212},
  year={2024}, 
doi={10.3233/FAIA240195}
}

@article{KasneciEtAl2023,
  author  = {Kasneci, Enkelejda and Se{\ss}ler, Kathrin and K{\"u}chemann, Stefan and Bannert, Maria and Dementieva, Daryna and Fischer, Frank and Gasser, Urs and Groh, Georg and G{\"u}nnemann, Stephan and H{\"u}llermeier, Eyke and Krusche, Stephan and Kutyniok, Gitta and Michaeli, Tilman and Nerdel, Claudia and Pfeffer, J{\"u}rgen and Poquet, Oleksandra and Sailer, Michael and Schmidt, Albrecht and Seidel, Tina and Stadler, Matthias and Weller, Jochen and Kuhn, Jochen and Kasneci, Gjergji},
  title   = {{ChatGPT for good? On opportunities and challenges of Large Language Models for education}},
  journal = {Learning and Individual Differences},
  year    = {2023},
  volume  = {103},
  pages   = {102274},
  doi     = {10.1016/j.lindif.2023.102274},
  url     = {https://doi.org/10.1016/j.lindif.2023.102274}
}

@article{DahlkemperEtAl2023,
  author  = {Dahlkemper, Merten Nikolay and Lahme, Simon Zacharias and Klein, Pascal},
  title   = {{How do physics students evaluate artificial intelligence responses on comprehension questions? A study on the perceived scientific accuracy and linguistic quality of {ChatGPT}}},
  journal = {Physical Review Physics Education Research},
  year    = {2023},
  volume  = {19},
  number  = {1},
  pages   = {010142},
  doi     = {10.1103/PhysRevPhysEducRes.19.010142},
  url     = {https://doi.org/10.1103/PhysRevPhysEducRes.19.010142}
}

@article{Renkl1997,
  author  = {Renkl, Alexander},
  title   = {Learning from Worked-Out Examples: A Study on Individual Differences},
  journal = {Cognitive Science},
  year    = {1997},
  volume  = {21},
  number  = {1},
  pages   = {1--29},
  doi     = {10.1207/s15516709cog2101_1},
  url     = {https://doi.org/10.1207/s15516709cog2101_1}
}

@article{GjerdeEtAl2022,
  author  = {Gjerde, Vegard and Paulsen, Vegard Havre and Holst, Bodil and Kolst{\o}, Stein Dankert},
  title   = {Problem solving in basic physics: Effective self-explanations based on four elements with support from retrieval practice},
  journal = {Physical Review Physics Education Research},
  year    = {2022},
  volume  = {18},
  number  = {1},
  pages   = {010136},
  doi     = {10.1103/PhysRevPhysEducRes.18.010136},
  url     = {https://doi.org/10.1103/PhysRevPhysEducRes.18.010136}
}

@article{BadeauEtAl2017,
  author  = {Badeau, Ryan and White, Daniel R. and Ibrahim, Bashirah and Ding, Lin and Heckler, Andrew F.},
  title   = {What works with worked examples: Extending self-explanation and analogical comparison to synthesis problems},
  journal = {Physical Review Physics Education Research},
  year    = {2017},
  volume  = {13},
  number  = {2},
  pages   = {020112},
  doi     = {10.1103/PhysRevPhysEducRes.13.020112},
  url     = {https://doi.org/10.1103/PhysRevPhysEducRes.13.020112}
}

@article{SchmidtWeigandEtAl2009,
  author  = {Schmidt-Weigand, Florian and H{\"a}nze, Martin and Wodzinski, Rita},
  title   = {Complex Problem Solving and Worked Examples: The Role of Prompting Strategic Behavior and Fading-in Solution Steps},
  journal = {Zeitschrift f{\"u}r P{\"a}dagogische Psychologie},
  year    = {2009},
  volume  = {23},
  number  = {2},
  pages   = {129--138},
  doi     = {10.1024/1010-0652.23.2.129},
  url     = {https://doi.org/10.1024/1010-0652.23.2.129}
}

@article{RenklAtkinson2003,
  author  = {Renkl, Alexander and Atkinson, Robert K.},
  title   = {Structuring the Transition From Example Study to Problem Solving in Cognitive Skill Acquisition: A Cognitive Load Perspective},
  journal = {Educational Psychologist},
  year    = {2003},
  volume  = {38},
  number  = {1},
  pages   = {15--22},
  doi     = {10.1207/S15326985EP3801_3}
}

@article{BarbieriEtAl2023,
  author  = {Barbieri, Christina Areizaga and Miller-Cotto, Dana and Clerjuste, Sarah N. and Chawla, Kamal},
  title   = {A Meta-analysis of the Worked Examples Effect on Mathematics Performance},
  journal = {Educational Psychology Review},
  year    = {2023},
  volume  = {35},
  number  = {1},
  pages   = {11},
  doi     = {10.1007/s10648-023-09745-1}
}

@article{davis1989,
  author  = {Davis, Fred D.},
  title   = {Perceived usefulness, perceived ease of use, and user acceptance of information technology},
  journal = {MIS Quarterly},
  year    = {1989},
  volume  = {13},
  number  = {3},
  pages   = {319--340},
  doi     = {10.2307/249008},
}

@article{miljkovic2024,
  author  = {Miljković Krečar, Irena and Kolega, Maja and Jurčec, Lana},
  title   = {Perception of {ChatGPT} usage for homework assignments: Students' and professors' perspectives},
  journal = {IAFOR Journal of Education: Technology in Education},
  year    = {2024},
  volume  = {12},
  number  = {2},
  pages   = {33--60},
doi={10.22492/ije.12.2.02}
}

@article{bitzenbauer2023,
  author  = {Bitzenbauer, Philipp},
  title   = {{ChatGPT} in physics education: {A} pilot study on easy-to-implement activities},
  journal = {Contemporary Educational Technology},
  year    = {2023},
  volume  = {15},
  number  = {3},
  pages   = {ep430},
  doi     = {10.30935/cedtech/13176},
}

@article{ChiWylie2014,
  author  = {Chi, Michelene T. H. and Wylie, Ruth},
  title   = {The {ICAP} Framework: Linking Cognitive Engagement to Active Learning Outcomes},
  journal = {Educational Psychologist},
  year    = {2014},
  volume  = {49},
  number  = {4},
  pages   = {219--243},
  doi     = {10.1080/00461520.2014.965823}
}

@article{BauerEtAl2025,
  author  = {Bauer, Elisabeth and Greiff, Samuel and Graesser, Arthur C. and Scheiter, Katharina and Sailer, Michael},
  title   = {Looking Beyond the Hype: Understanding the Effects of {AI} on Learning},
  journal = {Educational Psychology Review},
  year    = {2025},
  volume  = {37},
  pages   = {45},
  doi     = {10.1007/s10648-025-10020-8}
}

@article{PepperEtAl2012,
  author  = {Pepper, Rachel E. and Chasteen, Stephanie Viola and Pollock, Steven J. and Perkins, Katherine K.},
  title   = {Observations on student difficulties with mathematics in upper-division electricity and magnetism},
  journal = {Physical Review Special Topics - Physics Education Research},
  volume  = {8},
  number  = {1},
  pages   = {010111},
  year    = {2012},
  doi     = {10.1103/PhysRevSTPER.8.010111}
}

@article{CamposEtAl2023,
  author  = {Campos, Esmeralda and Hernandez, Eder and Barniol, Pablo and Zavala, Genaro},
  title   = {Analysis and comparison of students' conceptual understanding of symmetry arguments in {Gauss's} and {Ampere's} laws},
  journal = {Physical Review Physics Education Research},
  volume  = {19},
  number  = {1},
  pages   = {010103},
  year    = {2023},
  doi     = {10.1103/PhysRevPhysEducRes.19.010103}
}

@article{HernandezEtAl2023,
  author  = {Hernandez, Eder and Campos, Esmeralda and Barniol, Pablo and Zavala, Genaro},
  title   = {Students' conceptual understanding of electric flux and magnetic circulation},
  journal = {Physical Review Physics Education Research},
  volume  = {19},
  number  = {1},
  pages   = {013102},
  year    = {2023},
  doi     = {10.1103/PhysRevPhysEducRes.19.013102}
}

\end{document}